\documentclass[]{pasj01}
\onecolumn

\Received{}
\Accepted{}
 
\begin{document} 

\title{ 
On the influence of shock-cloud interactions on the nonthermal X-ray emission from the supernova remnant RCW 86}

\author{Aya \textsc{Bamba}\altaffilmark{1,2,3}}
\author{Hidetoshi \textsc{Sano}\altaffilmark{4}}
\author{Ryo \textsc{Yamazaki}\altaffilmark{5,6}}
\author{Jacco \textsc{Vink}\altaffilmark{7}}
\altaffiltext{1}{Department of Physics, Graduate School of Science,
The University of Tokyo, 7-3-1 Hongo, Bunkyo-ku, Tokyo 113-0033, Japan}
\email{bamba@phys.s.u-tokyo.ac.jp}
\altaffiltext{2}{Research Center for the Early Universe, School of Science, The University of Tokyo, 7-3-1
Hongo, Bunkyo-ku, Tokyo 113-0033, Japan}
\altaffiltext{3}{Trans-Scale Quantum Science Institute, The University of Tokyo, Tokyo  113-0033, Japan}
\altaffiltext{4}{Faculty of Engineering, Gifu University, 1-1 Yanagido, Gifu 501-1193, Japan}
\altaffiltext{5}{Department of Physical Sciences, Aoyama Gakuin University 5-10-1 Fuchinobe Chuo-ku, Sagamihara, Kanagawa 252-5258, Japan}
\altaffiltext{6}{Institute of Laser Engineering, Osaka University, 2-6 Yamadaoka, Suita, Osaka 565-0871, Japan}
\altaffiltext{7}{Anton Pannekoek Institute for Astronomy \& GRAPPA, University of Amsterdam, Science Park 904, 1098 XH Amsterdam, The
Netherlands}



\maketitle

\begin{abstract}
It is an open issue how the surrounding environment of supernova remnant shocks affect nonthermal X-rays from accelerated electrons,
with or without interacting dense material.
We have conducted spatially resolved X-ray spectroscopy of
the shock-cloud interacting region of RCW~86
with XMM-Newton.
It is found that bright soft X-ray filaments surround the dense cloud observed with $^{12}$CO and H$_{\rm I}$ emission lines.
These filaments are brighter in thermal X-ray emission, and fainter and possibly softer in synchrotron X-rays,
compared to those without interaction.
Our results show that the shock decelerates due to the interaction with clouds,
which 
results in an enhancements of thermal X-ray emission. This could possibly also explain the softer
X-ray synchrotron component,
because it implies
that
those shocks that move through a low density environment, and therefore decelerate much less,
can be more efficient accelerators.
This is similar to SN~1006 and Tycho, and is in contrast to RX~J1713.7$-$3946.
This difference among remnants may be due to the clumpiness
of dense material interacting with the shock,
which should be examined with future observations.
\end{abstract}


\section{Introduction}

Shocks of supernova remnants (SNRs) are the most plausible site of Galactic cosmic ray acceleration.
Together with X-ray observations,
it is known that 
the magnetic field on the shocks is amplified and turbulent,
which makes thin and bright synchrotron X-ray filaments or knots on the shock
\citep[for example]{bell2004,bamba2003,bamba2005,vink2003,uchiyama2007}.
One of the most important remaining problem is 
what kind of environment makes 
such a magnetic field amplification
--- when the maximum energy of accelerated electron is limited by
synchrotron cooling,
the maximum energy of accelerated electrons is proportional to the square of the shock velocity 
(e.g., \cite{aharonian1999,yamazaki2006,zirakashvili2007}),
implying that the low density environment makes the maximum energy larger since the shock remains fast with a low deceleration. 
In fact, many synchrotron X-ray dominated SNRs emit no or only faint thermal X-ray emission was detected
\citep{koyama1997,slane2001,bamba2001,yamaguchi2004,bamba2012}.
\citet{lopez2015} made the spatially resolved spectroscopy of Tycho with NuSTAR,
and showed that the high shock-speed regions have high energy cut-off of synchrotron X-rays.
On the other hand, 
\citet{inoue2009,inoue2012} suggested that the shock-cloud interaction amplifies
turbulent magnetic field.
Such clumpy clouds also affect the GeV-TeV spectra from supernova remnants and X-ray variability
\citep{celli2019}.
\citet{sano2013} discovered some molecular cloud clumps in RX~J1713.7$-$3946 surrounded by synchrotron X-ray filaments,
and \citet{sano2015} showed that the high interstellar gas density regions tend to have harder X-ray spectra in this remnant.
These results confirm 
the scenario that a clumpy medium enhances turbulence and hence facilitates X-ray synchrotron emission.
We need more samples to investigate what happens on the shock-cloud interacting regions in supernova remnants.

RCW~86 is one of the SNRs emitting 
synchrotron X-rays \citep{bamba2000,borkowski2001},
GeV gamma-rays \citep{lemoine2012,yuan2014},
and very high energy gamma-rays \citep{hess2018}.
An interesting characteristic of this SNR is that
it emits not only synchrotron X-rays but also thermal X-rays,
which makes this target unique among synchrotron X-ray dominated SNRs.
The ratio of thermal and synchrotron X-rays in RCW~86 is different from position to position \citep{broersen2014,tsubone2017},
especially on the eastern rim \citep{vink2006},
which can be due to the location dependence of shock wave velocity.
The shock velocity measured by proper motion also changes rapidly especially on the eastern side
\citep{yamaguchi2016}.
This can be connected to the fact that the south eastern part interacts with $^{12}$CO and/or H$_{\rm I}$ clouds \citep{sano2017,sano2019}.
These facts make this SNR ideal to study 
the effect of shock-cloud interaction on the particle acceleration.

In this paper, we study the spatially resolved spectroscopy
of RCW~86 shock-cloud interacting region,
together with X-ray and the CO map comparisons,
in order to understand how the interaction affects
particle acceleration.
Section~\ref{sec:obs} describes the data set as well as the data reduction.
The imaging and spectral analysis
results are described in section~\ref{sec:result}.
Finally, we discuss our results in section~\ref{sec:discuss}.
Throughout this paper,
we adopt 2.3~kpc as the distance to our target \citep{sollerman2003,helder2013}.

\section{Observations and Data reduction}
\label{sec:obs}

The southeast region of RCW86 was observed with XMM-Newton \citep{jansen2001} on 2014 January 27.
The data reduction and analysis was done with SAS version 20.0.0 \citep{gabriel2004}.
We selected only the data taken by the MOS cameras \citep{turner2001}
since the MOS camera has a better energy resolution and a lower background level compared with the pn camera.
We made the cleaned data with the standard method of XMM-Newton
following the SAS guide,
and the resultant exposure time is 101~ks.
For the spectral analysis, we used xspec 12.12.1 in headas 6.30.1.

\section{Results}
\label{sec:result}

\subsection{Images}

\begin{figure}
 \begin{center}
 \includegraphics[width=14cm]{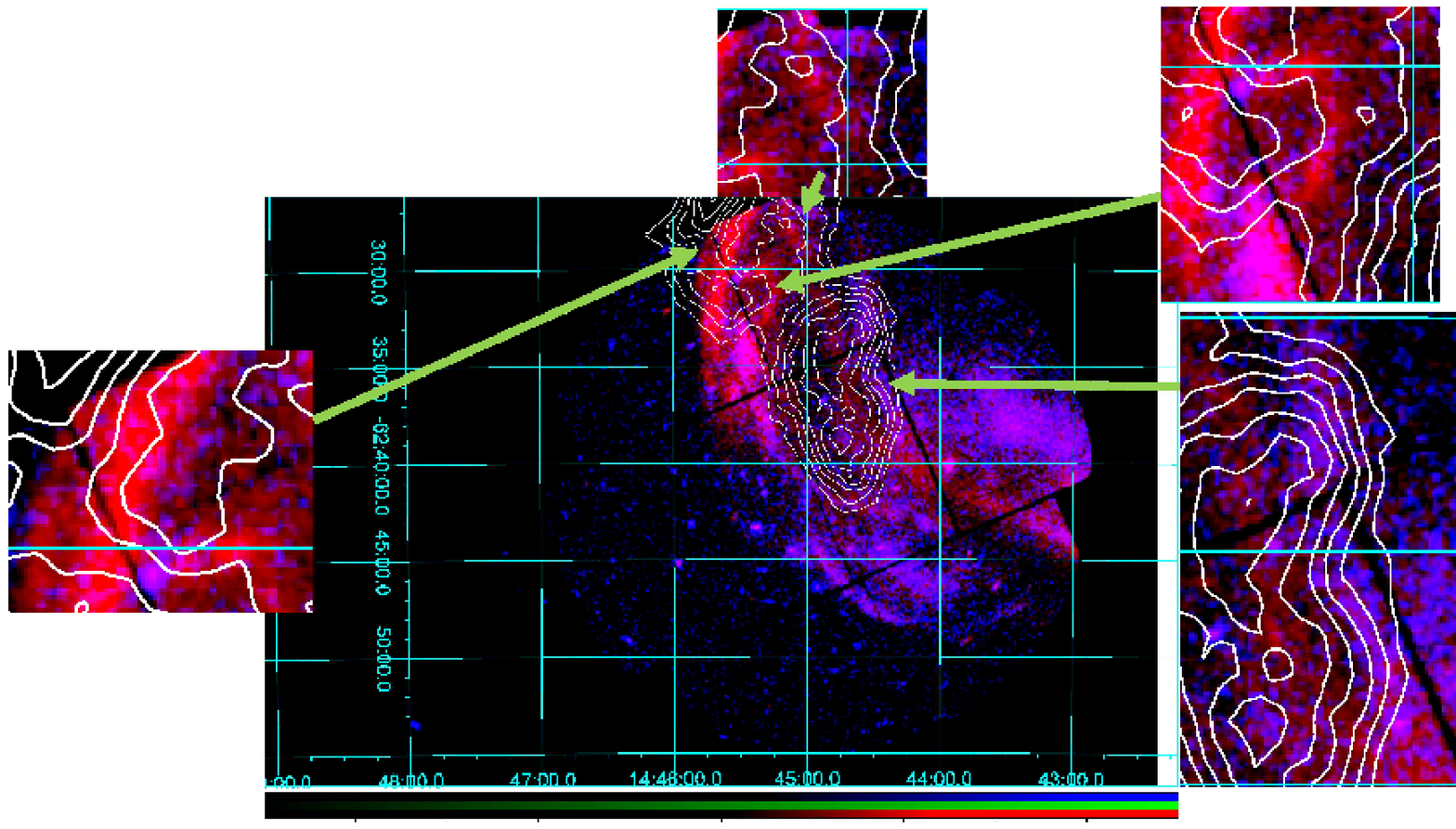}
 \end{center}
\caption{MOS2 0.5--2.0~keV (red) and 2.0--8.0~keV (blue) band images of 
the southeastern region of RCW~86 in the logarithmic scales.
The images are binned with 64$\times$64 pixels.
The vignetting correction and Non-X-ray-background subtraction is not performed.
Color range is between 0.2--20~cnt per binned pixel for the 0.5--2.0~keV band,
whereas 0.2--5 cnt per binned pixel.
Coordinates are in J2000.
White contours represent $^{12}$CO(J=2--1) distribution in the velocity range of $-$36 to $-$34~km~s$^{-1}$
taken by NANTEN2 (see Fig. 6 of \citet{sano2017}).
Small panels are closed up view of bright filaments.
}
\label{fig:image_CO}
\end{figure}

Figure~\ref{fig:image_CO} shows 
the MOS2 0.5--2.0~keV (red) and 2.0--8.0~keV (blue) image of the southeastern region of RCW~86.
We did not use MOS1 for the image analysis
since several chips are not in operation.
The $^{12}$CO(J=2--1) map is also overlaid with white contours.
One can see that several X-ray filaments beautifully surrounds
the molecular clouds
within a scale of $\sim$1~arcmin or $\sim$0.7~pc
at 2.3~kpc distance,
as if these filaments are aware of the contour levels, a result already reported on in  the original paper by \citet{sano2017}.

This result is already claimed by \citet{sano2017}.
The filaments surrounding the molecular clouds looks red,
implying that their X-ray emission is softer than
those from non-interacting regions.

\begin{figure}
 \begin{center}
\includegraphics[width=8cm]{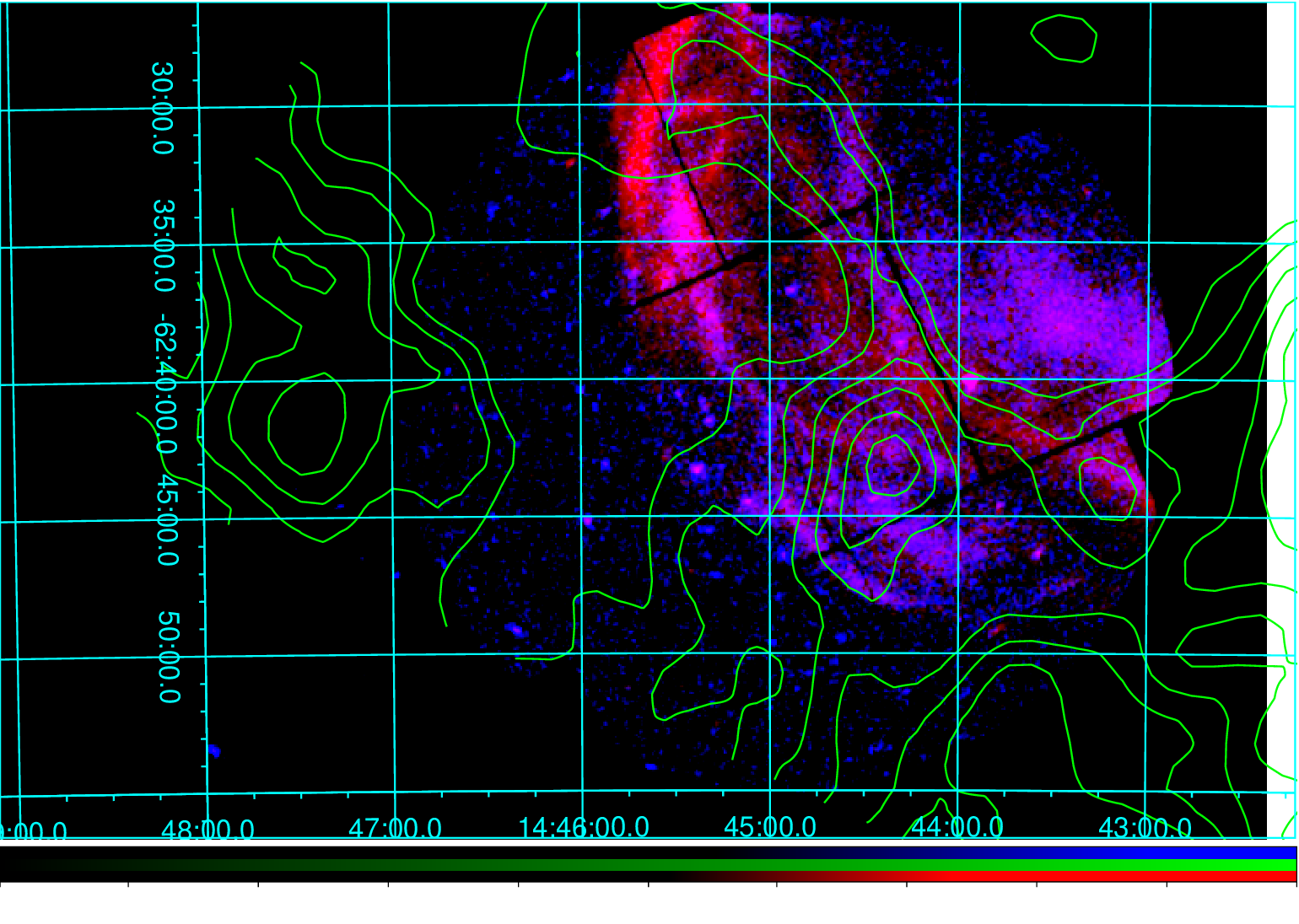}
 \end{center}
\caption{Same MOS2 images to Figure~\ref{fig:image_CO},
with H$_{\rm I}$ distribution in the velocity range of $-$35 to $-$30~km~s$^{-1}$ taken by ATCA \citep{sano2017}.
The contours start from 45~K in the linear scale, with the interval of 2.5~K.
}
\label{fig:image_HI}
\end{figure}

One can see that there are soft X-ray filaments extending southwest beyond the molecular cloud.
We thus examined whether there is dense material on these filaments
in the the Australia Telescope Compact Array (ATCA) H$_{\rm I}$ map taken by \citet{sano2017}.
Figure~\ref{fig:image_HI} shows the same MOS2 map with H$_{\rm I}$ contours.
It is found that there is an associated H$_{\rm I}$ cloud
in the velocity range between $-$35 and $-$30~km~s$^{-1}$,
which is similar in range to the velocity of the $^{12}$CO clouds that are associated with the SNR.
The filaments are on the edge of the H$_{\rm I}$ cloud,
which is the same situation as the northern CO cloud cases.

\subsection{Spectra}

\begin{figure}
 \begin{center}
\includegraphics[width=8cm]{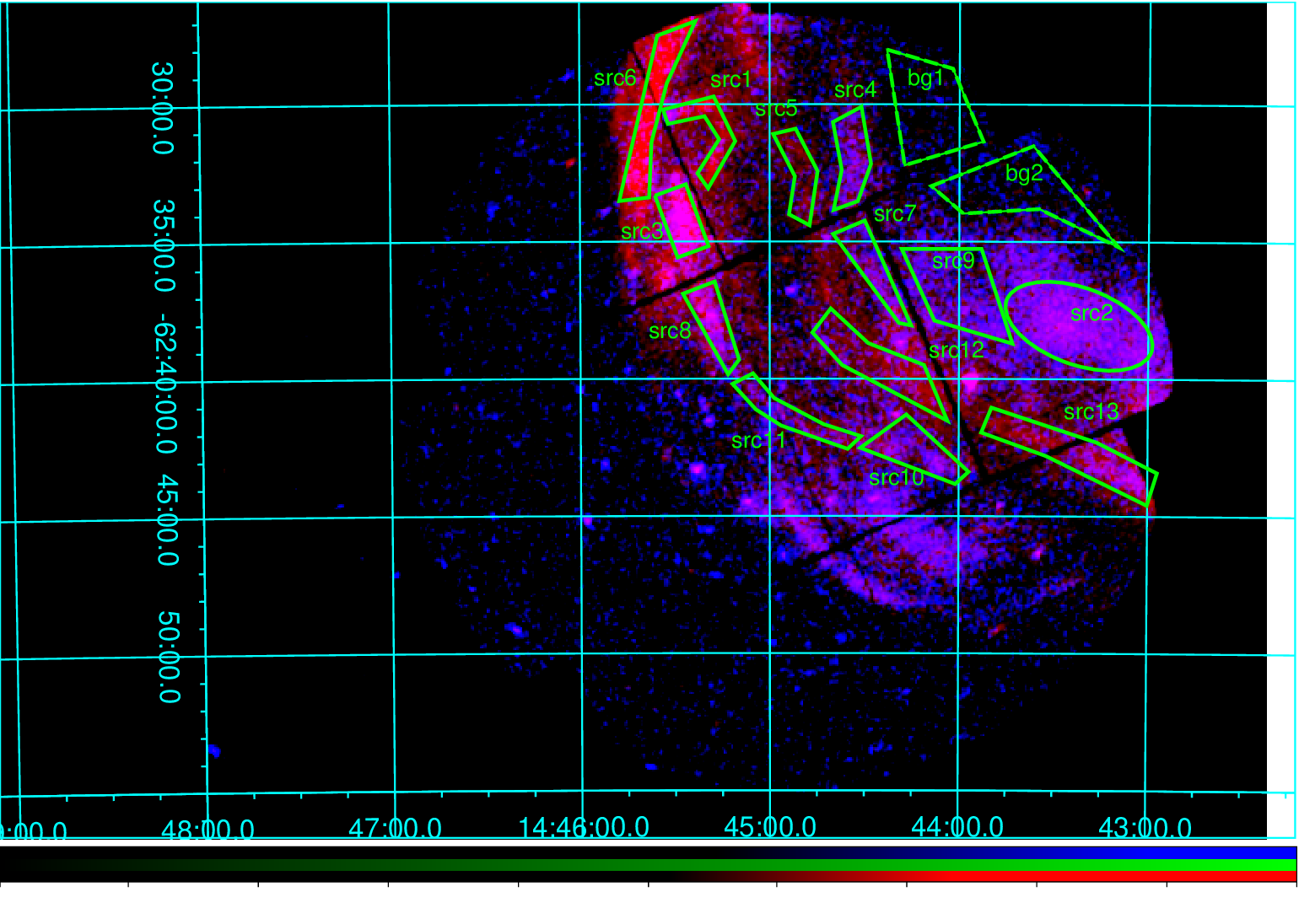}
 \end{center}
\caption{Same MOS2 images to Figure~\ref{fig:image_CO},
with source and background (solid and dashed, shown in green) regions for the spectral analysis.}
\label{fig:image_regions}
\end{figure}

In order to examine the spectral difference among filaments,
we selected thirteen regions for the spectral analysis,
as shown in Fig.~\ref{fig:image_regions}.
The background regions were selected as the dashed regions.
Since several regions are located outside of MOS1 field of view due to the lack of a CCD chip, we used only MOS2 camera data for such regions (src 1, 3, 4, 5, and 6).

\begin{figure}
 \begin{center}
\includegraphics[width=0.3\textwidth]{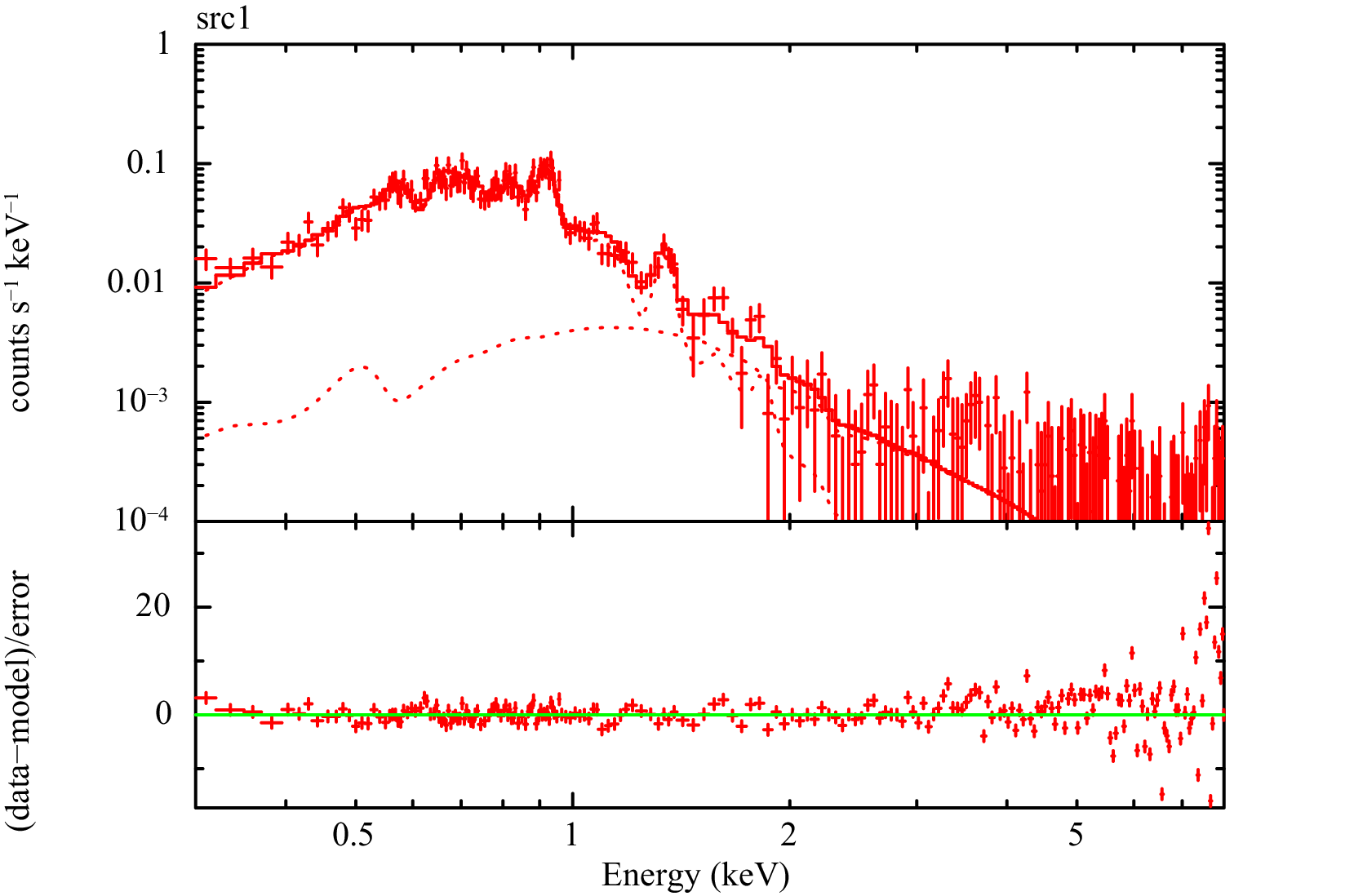}
\includegraphics[width=0.3\textwidth]{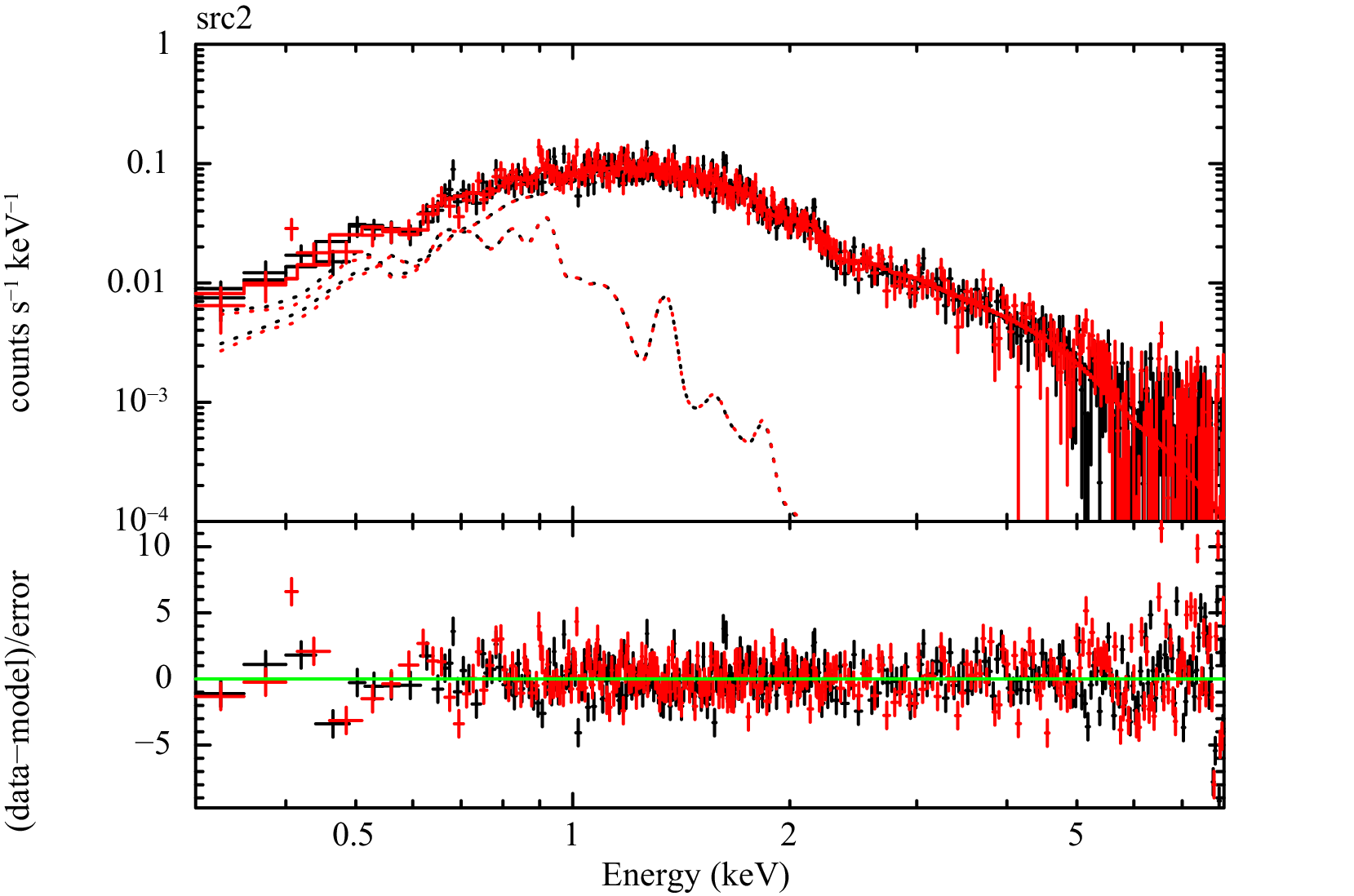}
\includegraphics[width=0.3\textwidth]{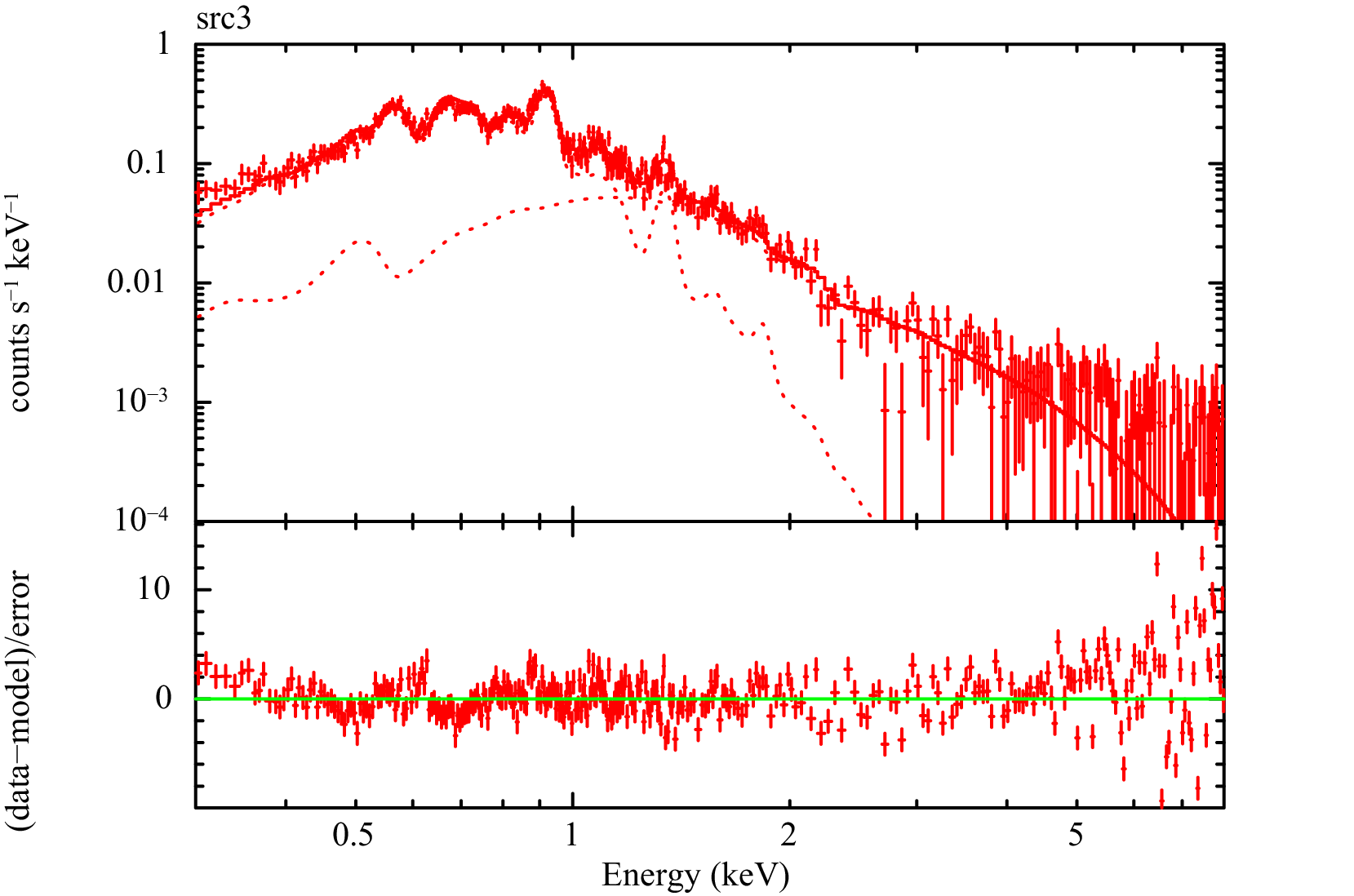}
\includegraphics[width=0.3\textwidth]{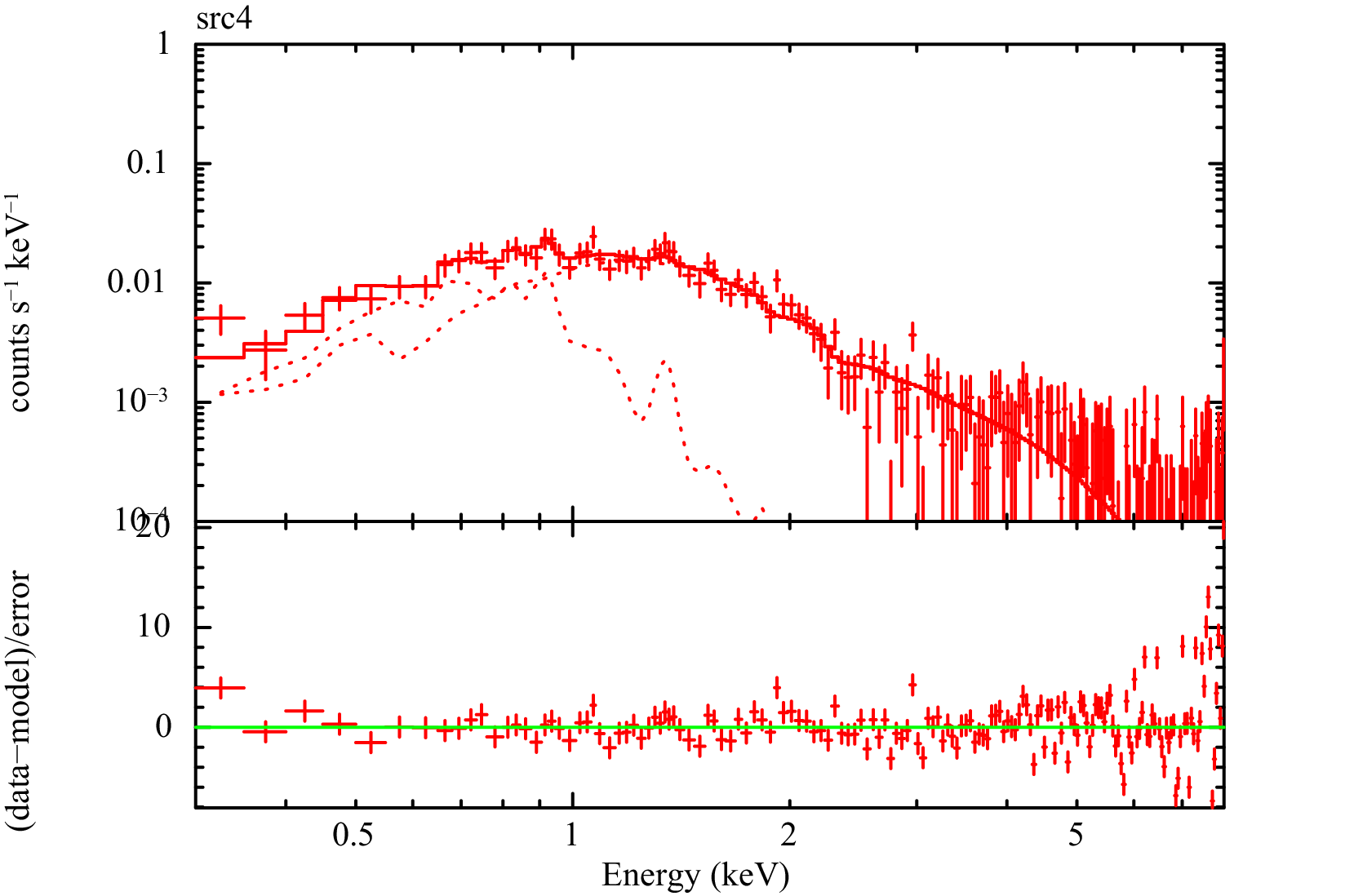}
\includegraphics[width=0.3\textwidth]{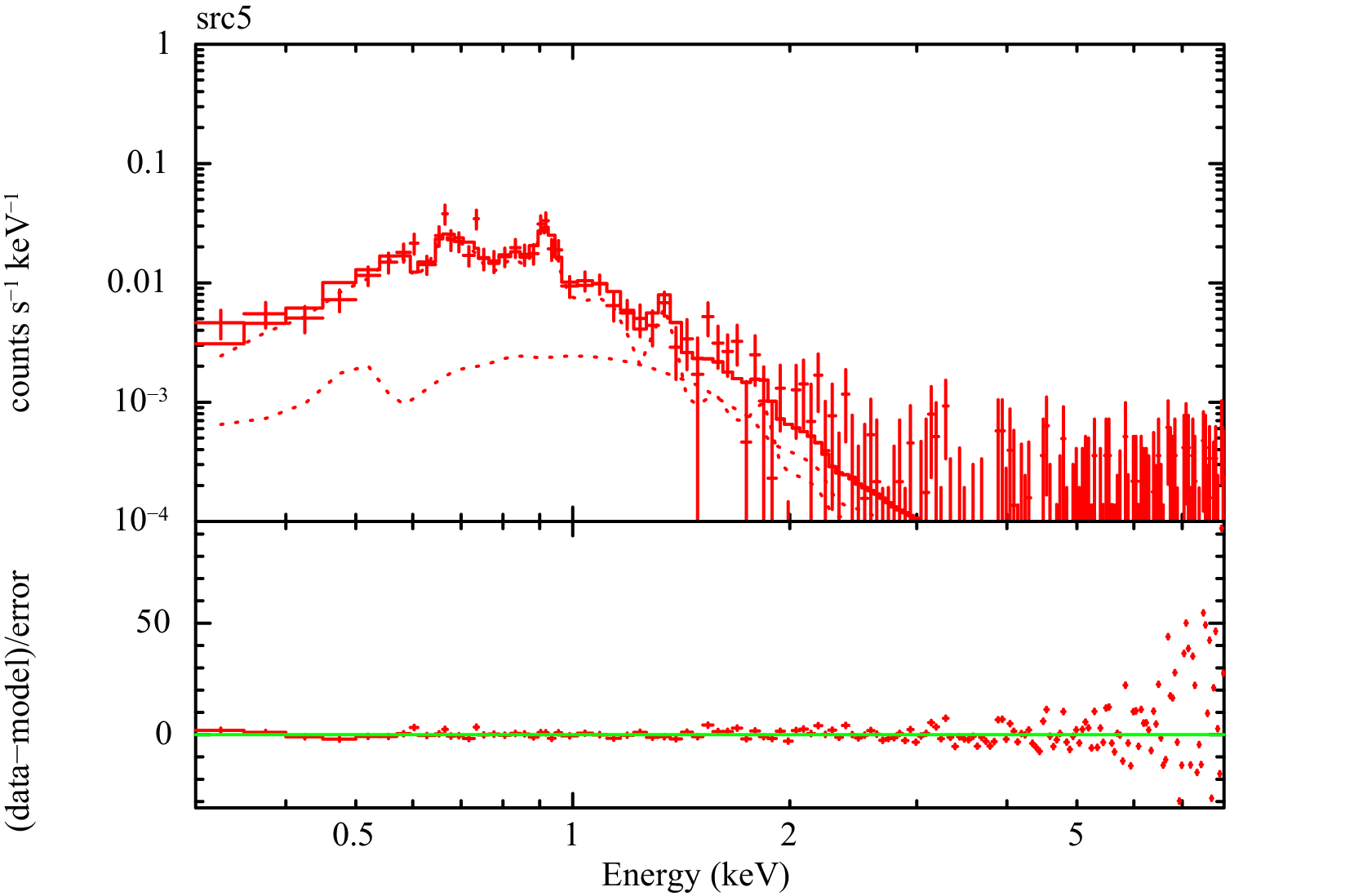}
\includegraphics[width=0.3\textwidth]{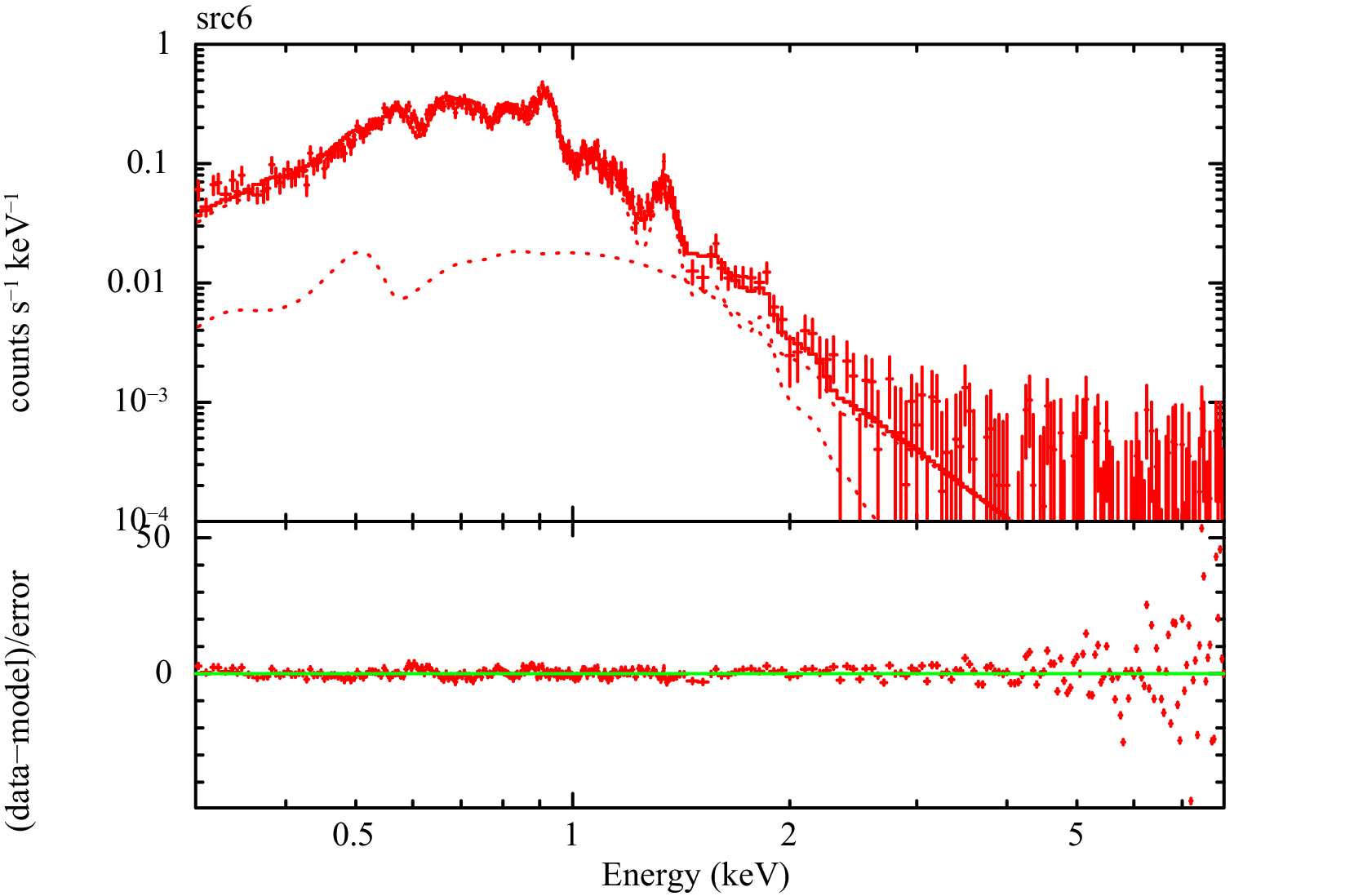}
\includegraphics[width=0.3\textwidth]{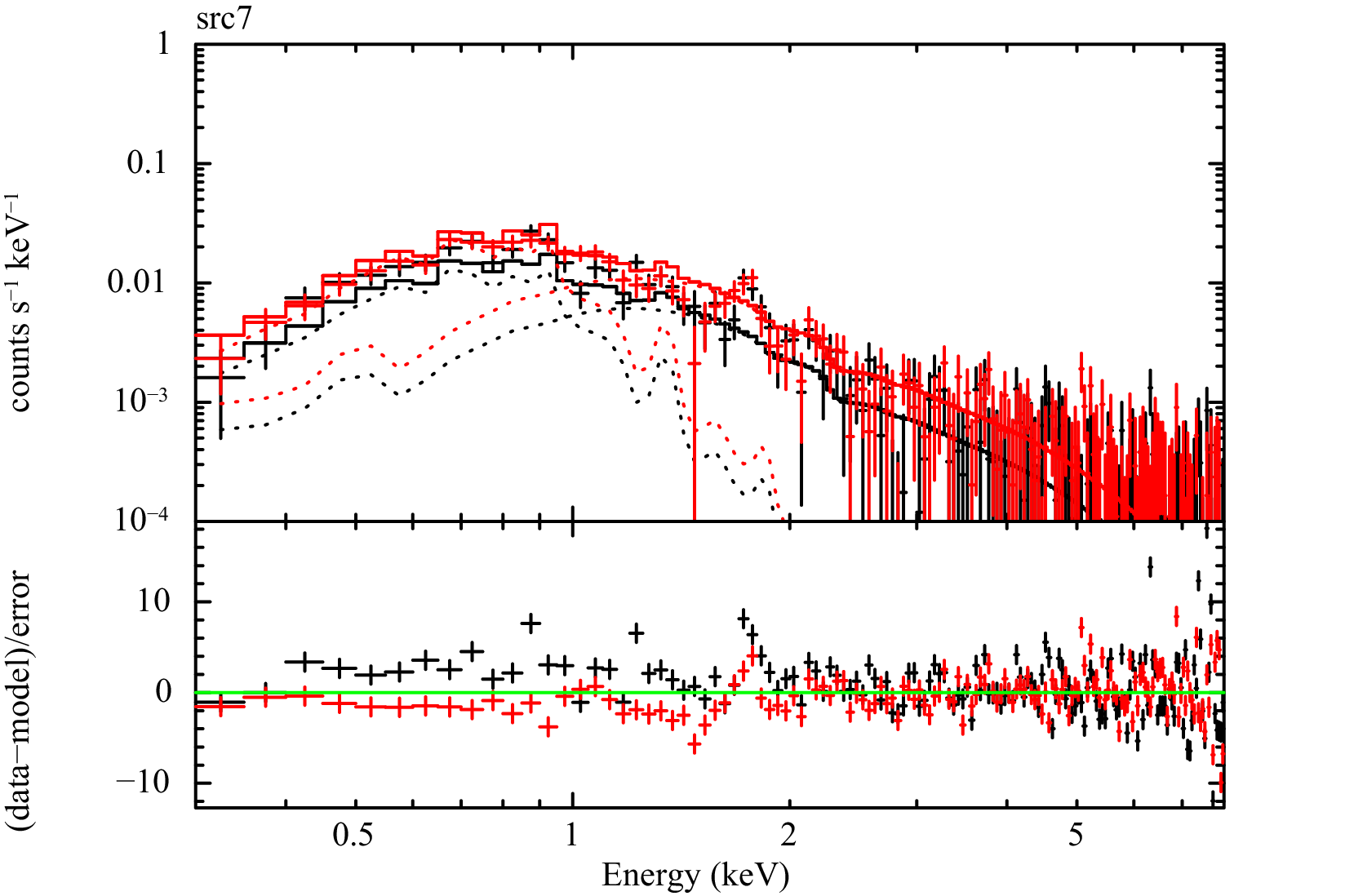}
\includegraphics[width=0.3\textwidth]{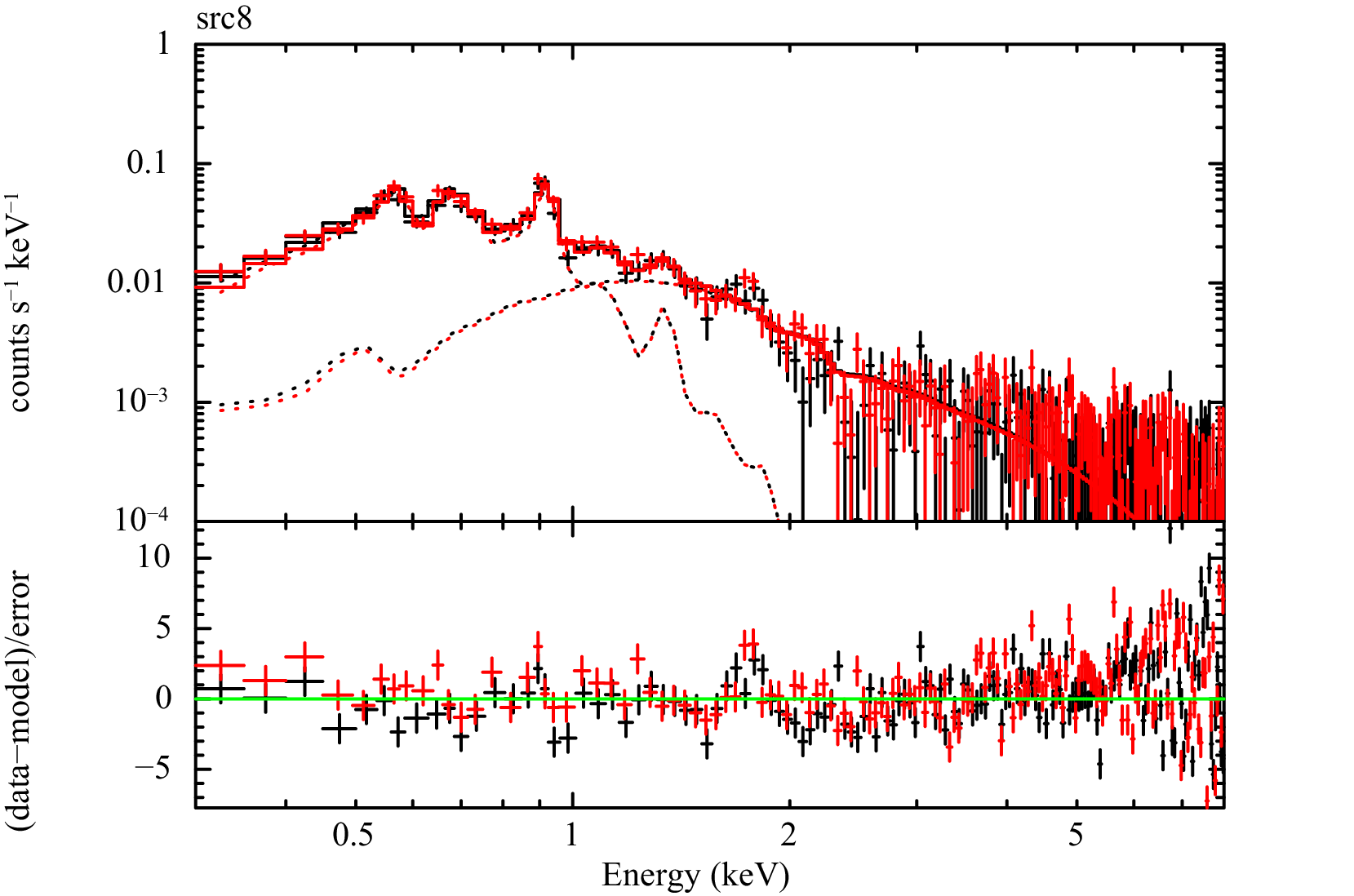}
\includegraphics[width=0.3\textwidth]{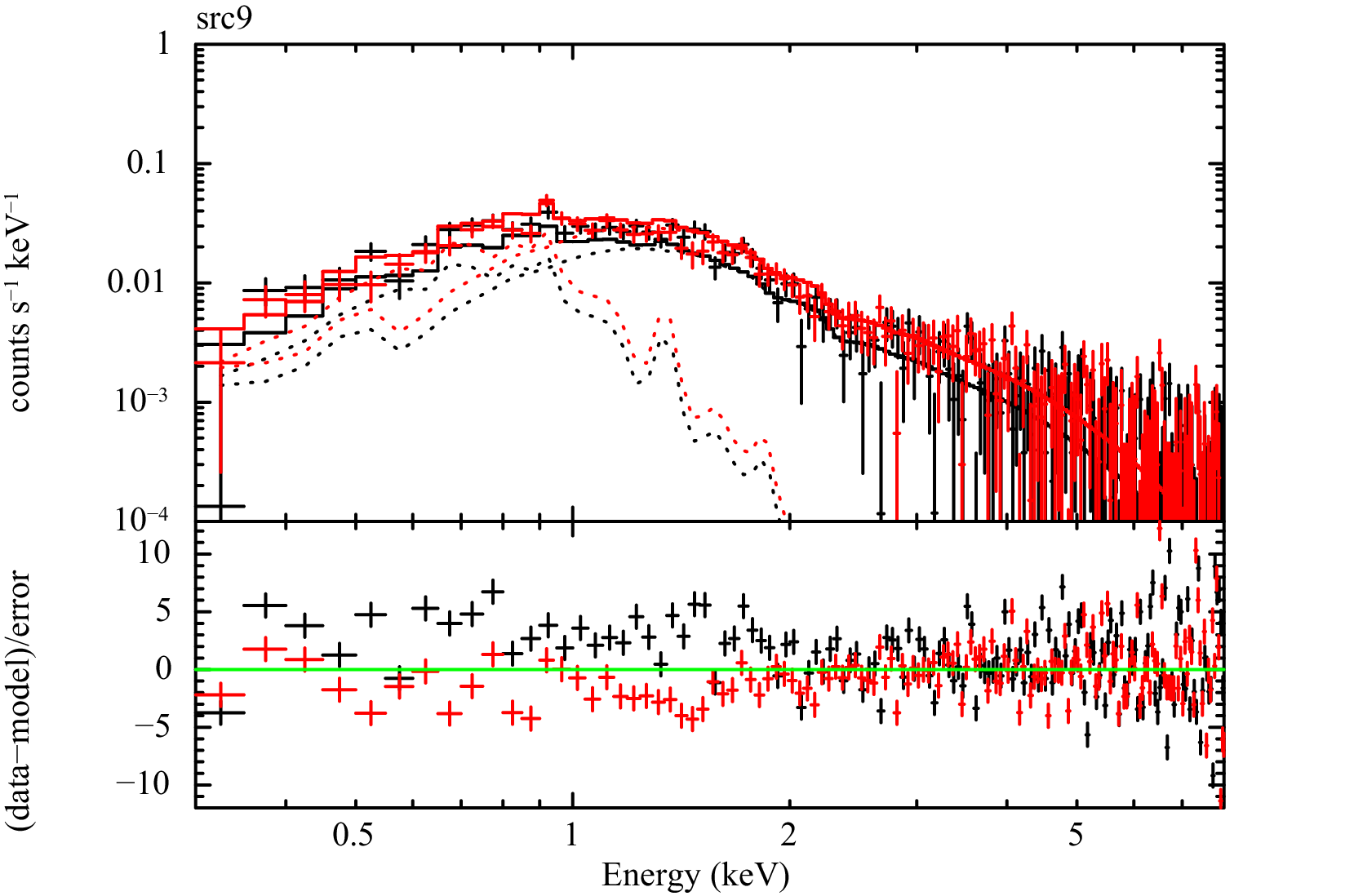}
\includegraphics[width=0.3\textwidth]{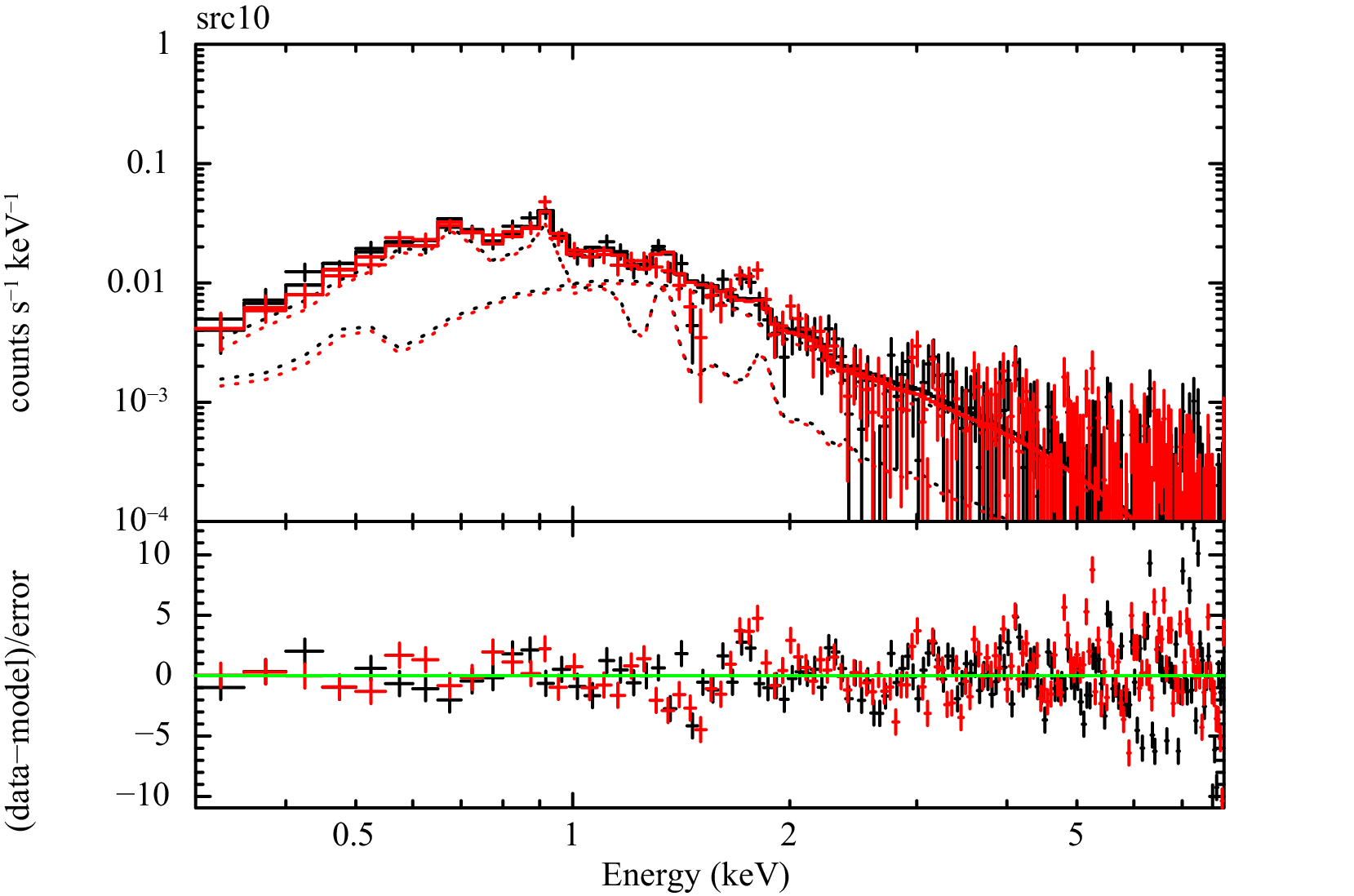}
\includegraphics[width=0.3\textwidth]{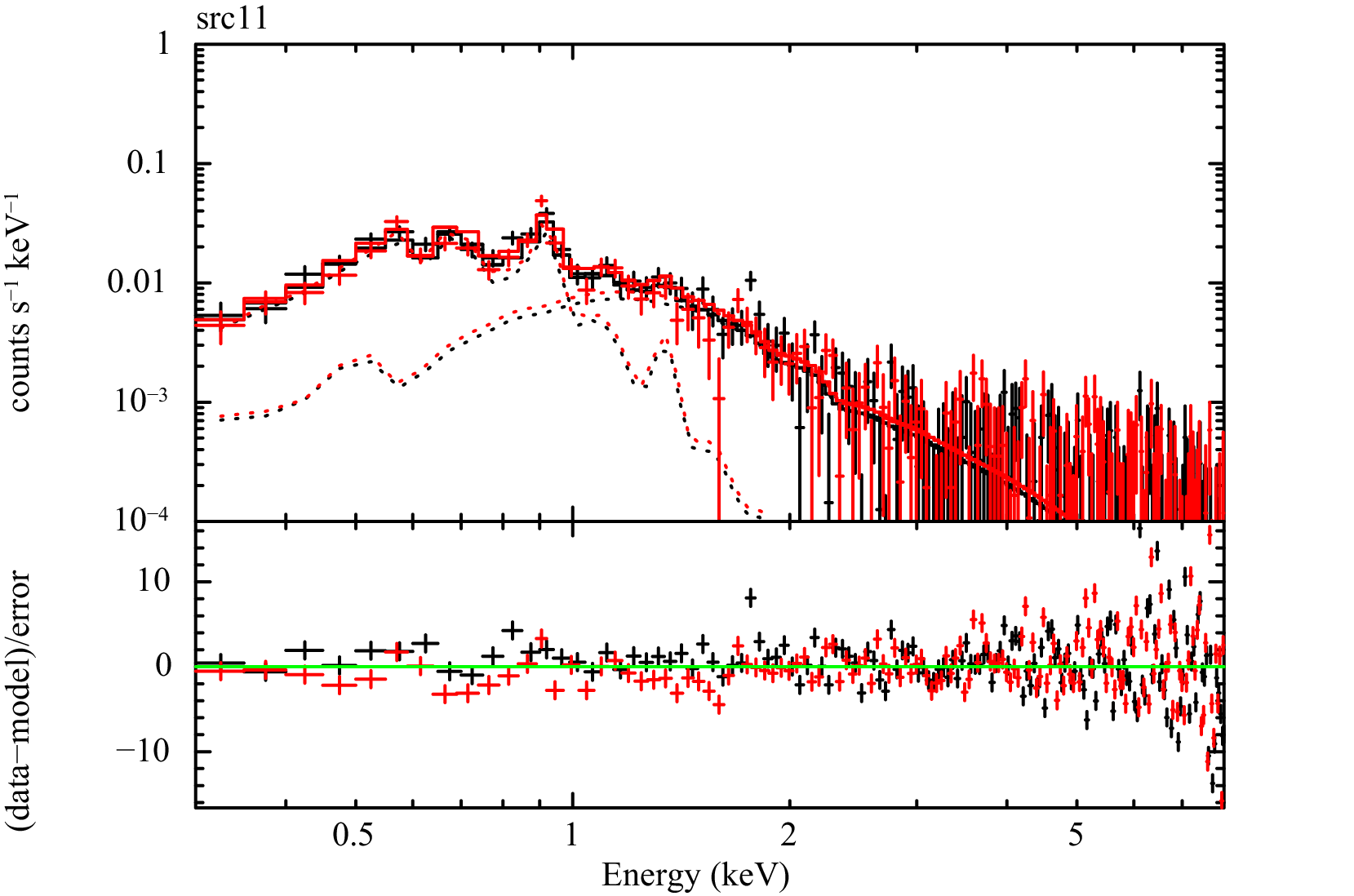}
\includegraphics[width=0.3\textwidth]{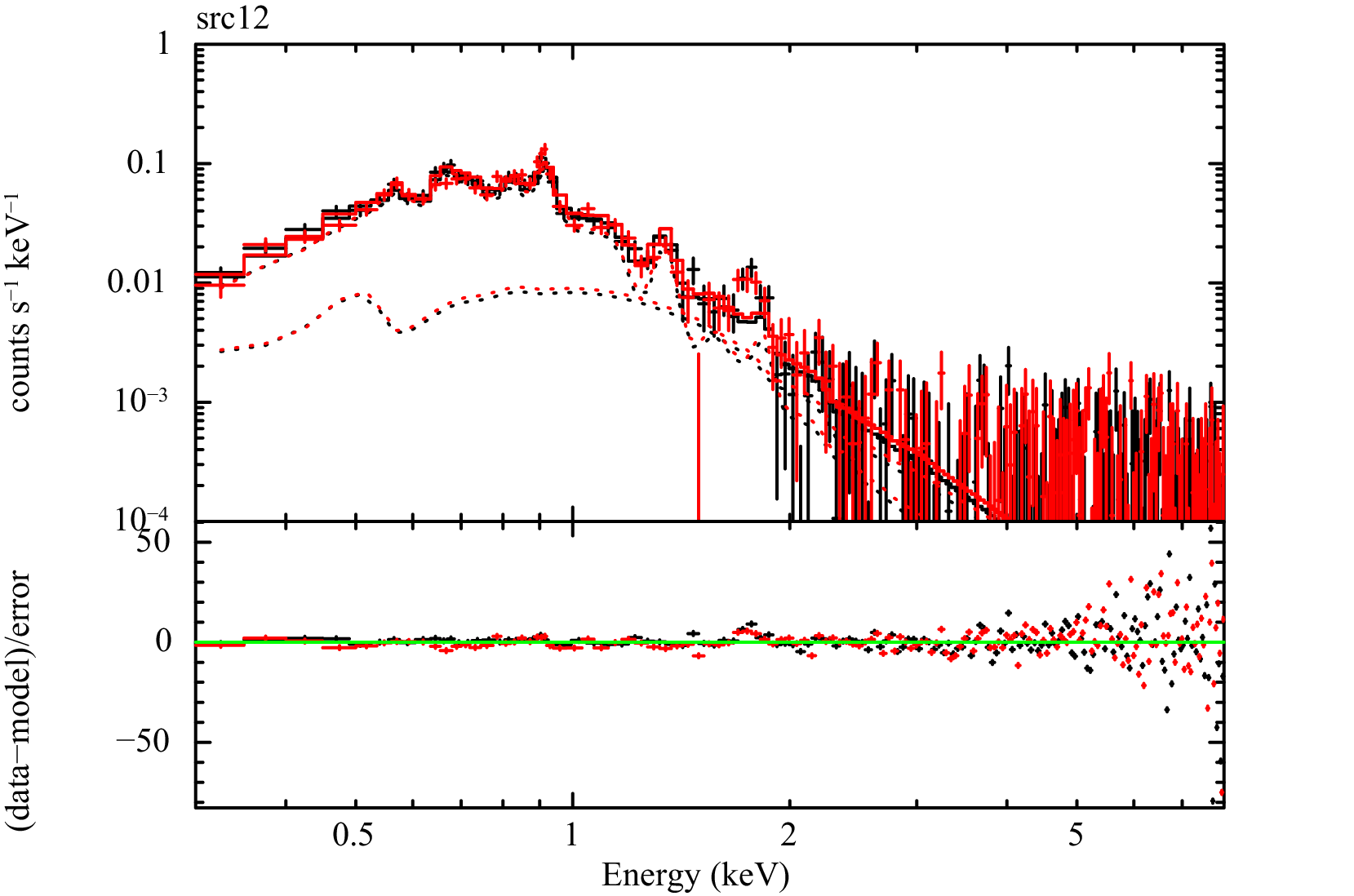}
\includegraphics[width=0.3\textwidth]{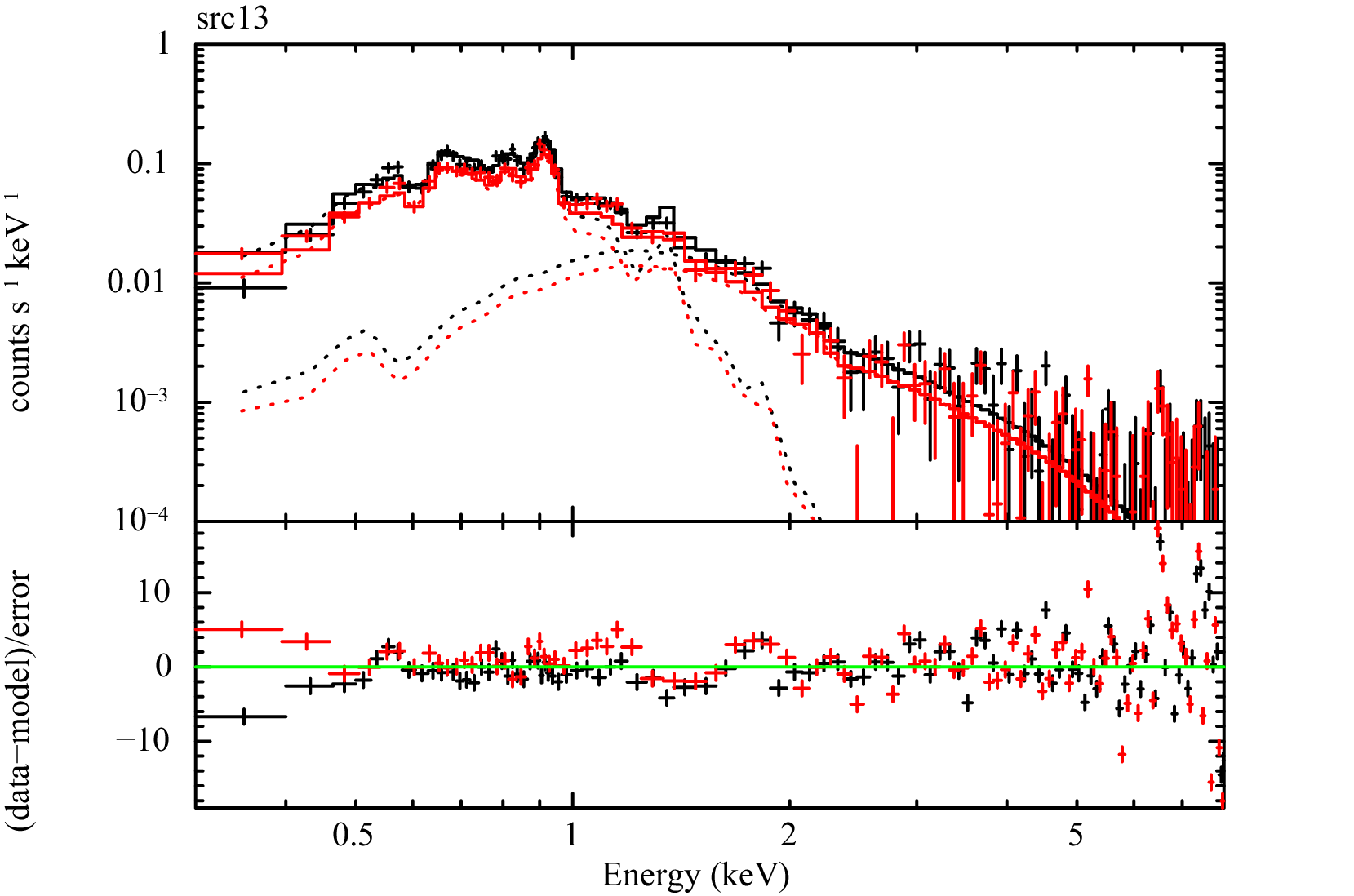}
 \end{center}
\caption{Spatially resolved spectra with best-fit models.
Black and red shows MOS1 and MOS2 data, respectively.
The dotted lines represent {\tt vnei} and power-law components.
The data is binned for better visibility,
whereas the fit is done without binning.}\label{fig:spectra}
\end{figure}

Figure~\ref{fig:spectra} show the background-subtracted spectrum
for each region.
One can see that some show softer spectra with emission lines, 
and other show harder spectra without lines.
We thus fitted all spectra with an non-equilibrium thermal component ({\tt nei} in xspec;  \cite{borkowski2001b}) plus a power-law component to reproduce synchrotron X-rays.
All the abundances are set to solar for the thermal emission.
To represent the absorption model,
we use the {\tt phabs} model in XSPEC,
which includes the cross sections of \citet{balucinska-church1992}
with solar abundances \citep{anders1989}.
We used Cash statistics for the spectral fitting
\citep{cash1979}.
The data are all well reproduced by these models,
as shown in Table~\ref{tab:fitting}.

Ideally we would have preferred to take larger background regions
in order to improve the statistics of the background spectra.
There exist larger source-free regions that could, in principle, be used for background subtraction.
However, we checked these regions and found that
the non- X-ray background (NXB) and the Galactic Ridge X-ray Emission (GRXE) varies a lot from position to position,
and these regions are, therefore, not suitable background spectra.
In addition we checked blank-sky data in the same detector area as the source regions, but here the difficulty is that the blank-sky regions are from high Galactic latitude regions and do not correctly predict the GRXE component.
For those reasons we 
prefer
the less ideal background spectra taken from close to the source regions. 
Unfortunately,
this does result in some residuals around the Si-K line band, most likely due to incorrectly taking into account the NXB components (src12 in Fig. 4 for example).  
Ignoring the Si-K line band for our fits produces slightly better fit statistics,
but the best-fit parameters were not affected 
significantly excluding this band, 
and, in fact, for those spectra without large residuals around Si-K the fit statistics
became even worse.
So in the end we kept the Si-K band in our fit results.
For src7 and 9, the spectra from MOS1 and MOS2 shows some normalization discrepancy.
It can be due to the fact that these regions are near the edge of CCDs and the calibration uncertainty of normalization can be large.
We thus made a simultaneous fit with coupled parameters but different normalization between MOS1 and MOS2 for src7 and 9 spectra.
The best-fit normalization changed 20--30\%, but still within the error region of the original fitting,
because these regions are rather small and statistics are not so good.
Thus, for simplicity, we adopted the same fitting method
to other regions.

\begin{table}
  \tbl{Best-fit parameters for the spectral fitting.\footnotemark[$*$] }{%
  \begin{tabular}{cccccccccc}
      \hline
      Reg. & Area & $N_{\rm H}$ & $\Gamma$ & $F_{\rm 2-10keV}$\footnotemark[$\dag$] & $kT_{\rm e}$ & $n_{\rm e}t$ & Norm\footnotemark[$\S$] &  $F_{\rm 0.5-10keV}$\footnotemark[$\dag$]
    & C-stat./d.o.f.\\ 
      & (arcsec$^2$) & ($10^{21}$cm$^{-2}$) & 
 & & (keV) & ($10^{10}$~s~cm$^{-3}$) & \\
      \hline
src1 & 9349 & 4.1 & 3.11 & 0.17 & 0.43 & 1.4 & 30 & 23.5 & 1385.9/1533\\
 & & (3.7--5.1) & (2.45--3.85) & (0.09--0.28) & (0.25--0.57) & (1.0--4.2) & (17--135) & (17.3--36.3) \\
src2 & 44411 & 5.5 & 2.84 & 1.54 & ---\footnotemark[$\|$] & ---\footnotemark[$\|$] & 7.5 & 4.5 & 3339.3/3074\\
 & & (5.1--5.9) & (2.79--2.91) & (1.48--1.59) & & & (5.1--10.4) & (3.1--6.3) \\
src3 & 10373 & 4.6 & 3.31 & 0.28 & 0.34 & 1.6 & 36 & 20.9 & 1736.3/1533 \\
 & & (4.4--4.9) & (3.16--3.45) & (0.25--0.31) & (0.27--0.39) & (1.3--2.2) & (25--45) & (17.3--25.4) \\
src4 & 12114 & 5.6 & 3.25 & 0.41 & 0.24 & 6.2 & 16 & 4.8 & 1408.3/1533 \\
 & & (3.6--6.8) & (2.85--3.49) & (0.35--0.47) & (0.18--4.78) & (0.4--28.6) & (0.9--72) & (0.7--12.1) \\
src5 & 9902 & 4.3 & 4.16 & 0.02 & 0.66 & 0.7 & 4.4 & 5.0 & 1269.9/1533 \\
& & (3.7--5.2) & (3.38--5.07) & (0.008--0.05) & (0.29--1.01) & (0.5--2.2) & (2.5--17.0) & (3.4--10.3) \\
src6 & 18934 & 4.5 & 4.49 & 0.14 & 0.34 & 2.0 & 171 & 104.0 & 1564.0/1533 \\
 & & (4.1--4.7) & (4.10--4.85) &  (0.06--0.18) & (0.32--0.45) & (1.3--2.3) & (94--190) & (76.9--108.9) \\
src7 & 11448 & 4.6 & 2.83 & 0.32 & 0.35 & 2.1 & 7.3 & 4.6 & 3042.2/3072 \\
  & & (3.6--5.5) & (2.58--3.07) &  (0.28--0.37) & (0.24--0.95) & (0.8--6.3) &  (1.5--28) & (2.2--8.9) \\
src8 & 10301 & 4.8 & 2.86 & 0.38 & 0.24 & 2.5 & 57 & 18.8 & 3103.2/3072 \\
  & & (4.1--5.9) & (2.69--3.09) & (0.33--0.43) & (0.18--0.35) &  (1.1--10.6) &  (19--240) & (10.8--38.2) \\
src9 & 28590 & 5.4 & 2.92 & 1.22 & 0.34 & 2.2 & 15 & 9.1 & 3384.6/3072 \\
 & & (4.5--6.4) & (2.74--3.10) & (1.11--1.34) & (0.22--0.75) & (0.7--11.4) & (3.4--73) & (4.0--19.9) \\
src10 & 14664 & 3.8 & 2.85 & 0.39 & 1.88 & 0.42 & 2.6 & 4.4 & 2981.6/3072 \\
  & & (3.4--4.4) & (2.63--3.11) & (0.31--0.49) & (0.35--3.53) & (0.36--0.96) & (1.9--6.4) & (3.4--19.8) \\
src11 & 11786 & 5.5 & 3.50 & 0.13 & 0.21 & 4.1 & 50 & 11.9 & 3316.4/3078 \\
 & & (4.9--6.4) & (3.24--3.71) & (0.11--0.16) & (0.17--0.28) & (1.8--16.3) & (18--163) & (7.1--21.6) \\
src12 & 21340 & 4.0 & 4.13 & 0.043 & 0.69 & 0.8 & 6.8 & 8.0 &  3076.3/3072 \\
 & & (3.7--4.4) & (3.79--4.65) & (0.02--0.06) & (0.61--1.01) & (0.6--1.0) & (4.7--9.4) & (6.3--9.6) \\
src13 & 26501 & 6.4 & 3.36 & 0.50 & 0.20 & 14.1 & 417 & 85.9 &  2917.9/3078 \\
 & & (6.2--6.6) & (3.17--3.55) & (0.43--0.57) & (0.19--0.22) & (9.1--21.8) & (284--512) & (76.5--99.1) \\
 \hline
    \end{tabular}}\label{tab:fitting}
\begin{tabnote}
\footnotemark[$*$] The errors are the 90\% confidence level.  \\ 
\footnotemark[$\dag$] 2--10~keV flux of the power-law component in the unit of $10^{-12}$erg~cm$^{-2}$s$^{-1}$. \\
\footnotemark[$\ddag$] $\frac{10^{-18}}{4\pi D^2}\int n_en_H dV$, where $D$, $n_e$, and $n_H$ is the angular diameter distance to the source (cm), electron and hydrogen density (cm$^{-3}$), respectively.\\
\footnotemark[$\S$] 0.5--10~keV flux of the thermal component in the unit of $10^{-12}$erg~cm$^{-2}$s$^{-1}$. \\
\footnotemark[$\|$]  Fixed to the parameters for src9.\\
\end{tabnote}
\end{table}

\begin{figure}
 \begin{center}
\includegraphics[width=0.4\textwidth]{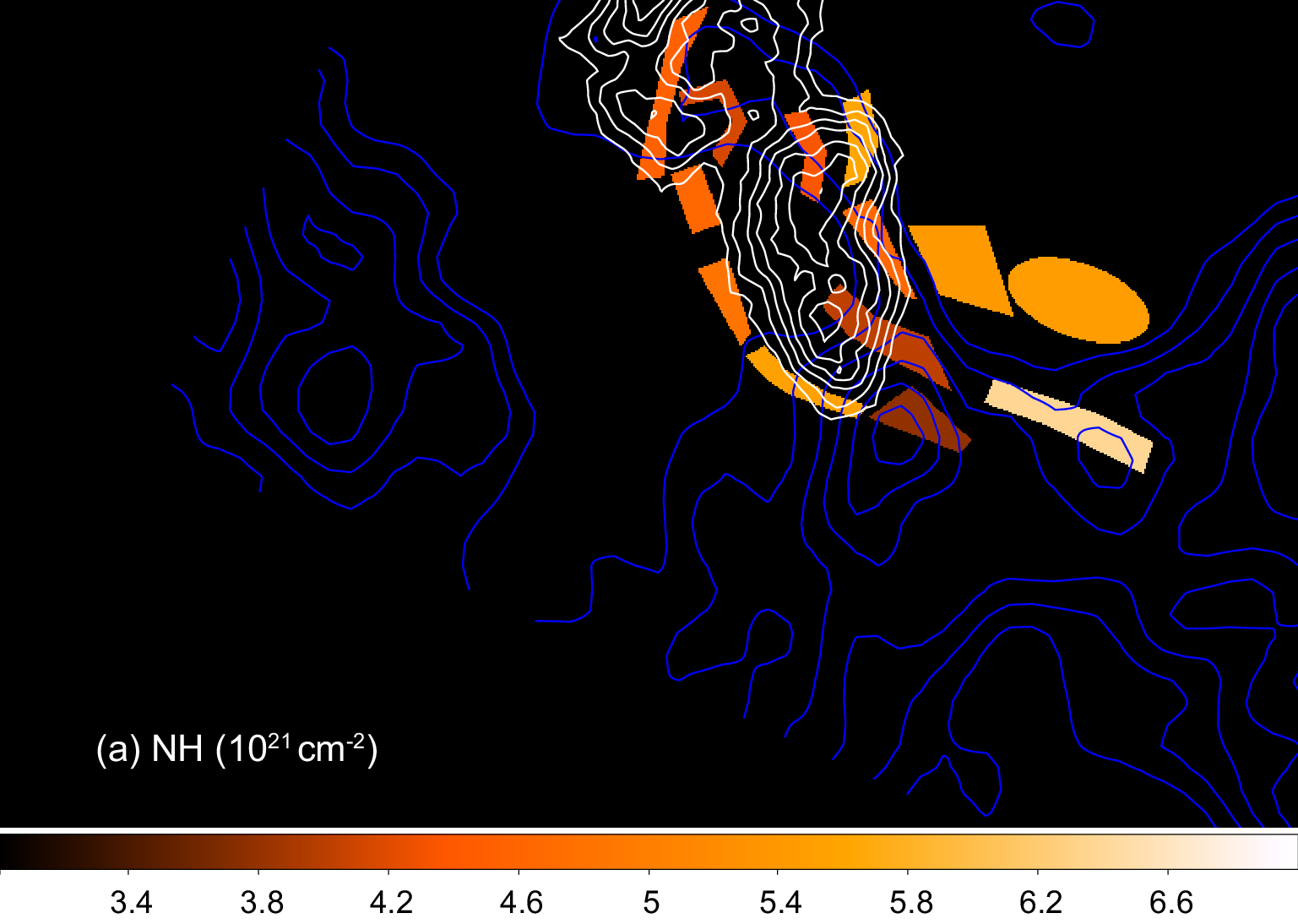}
\includegraphics[width=0.4\textwidth]{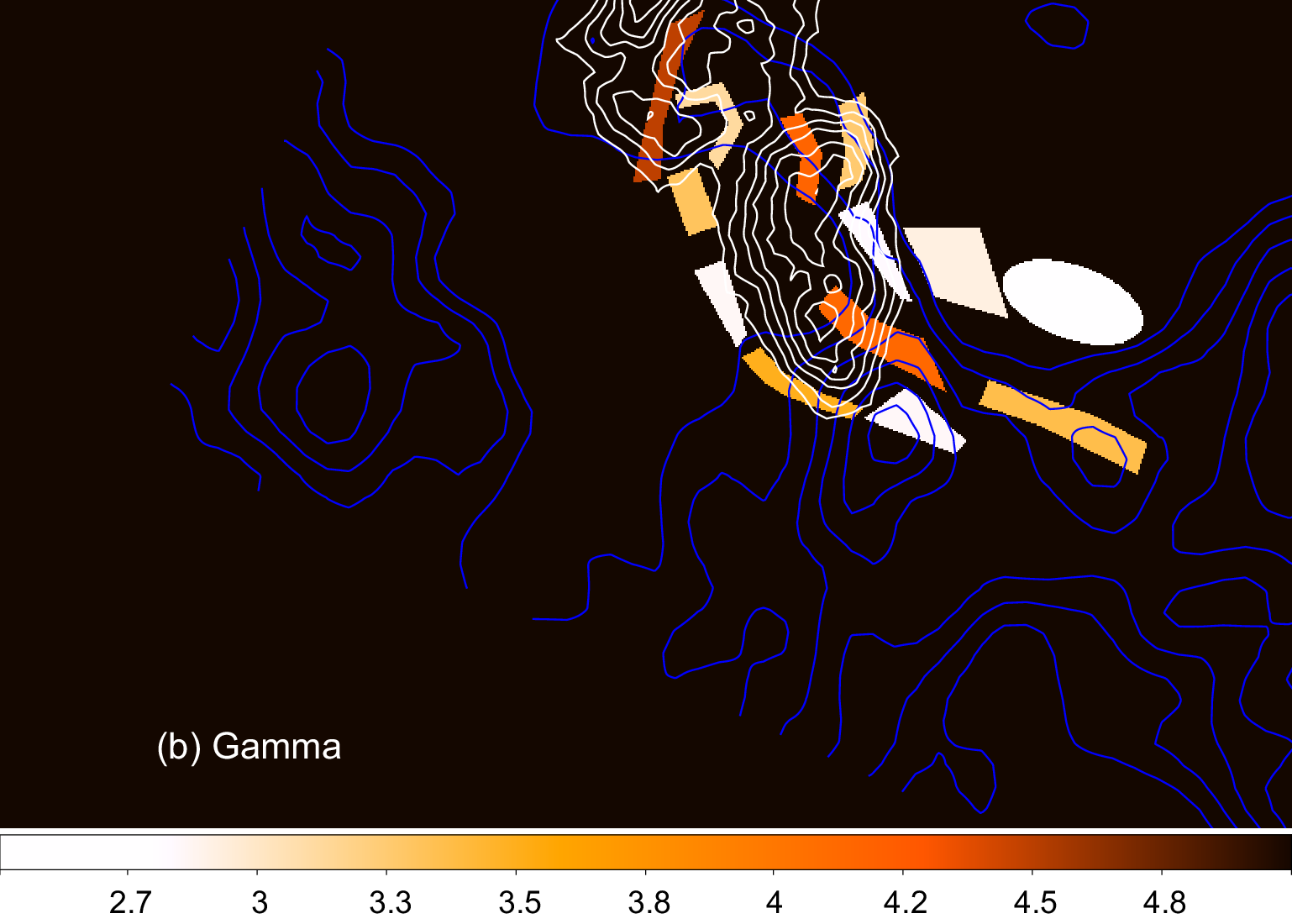}
\includegraphics[width=0.4\textwidth]{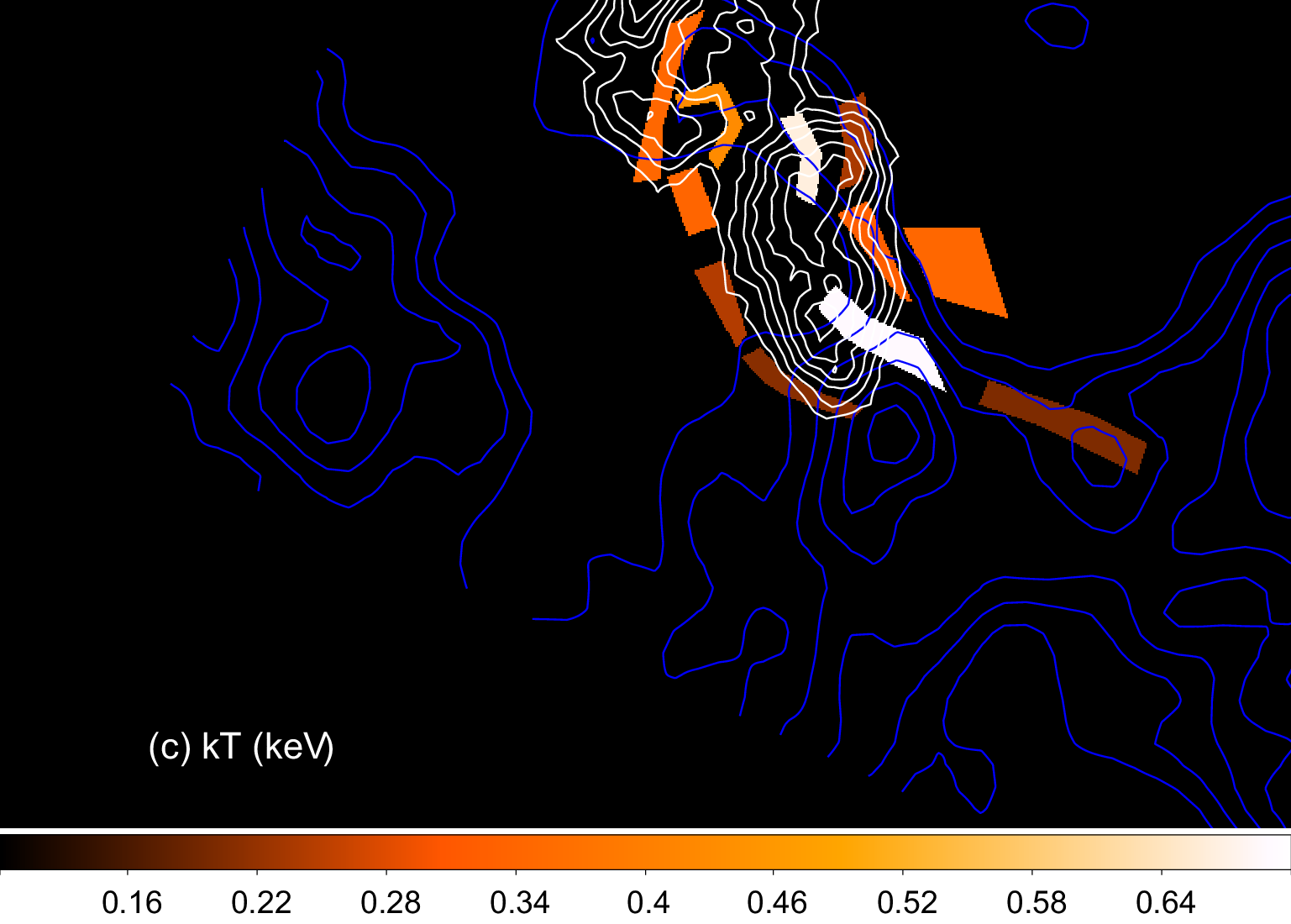}
\includegraphics[width=0.4\textwidth]{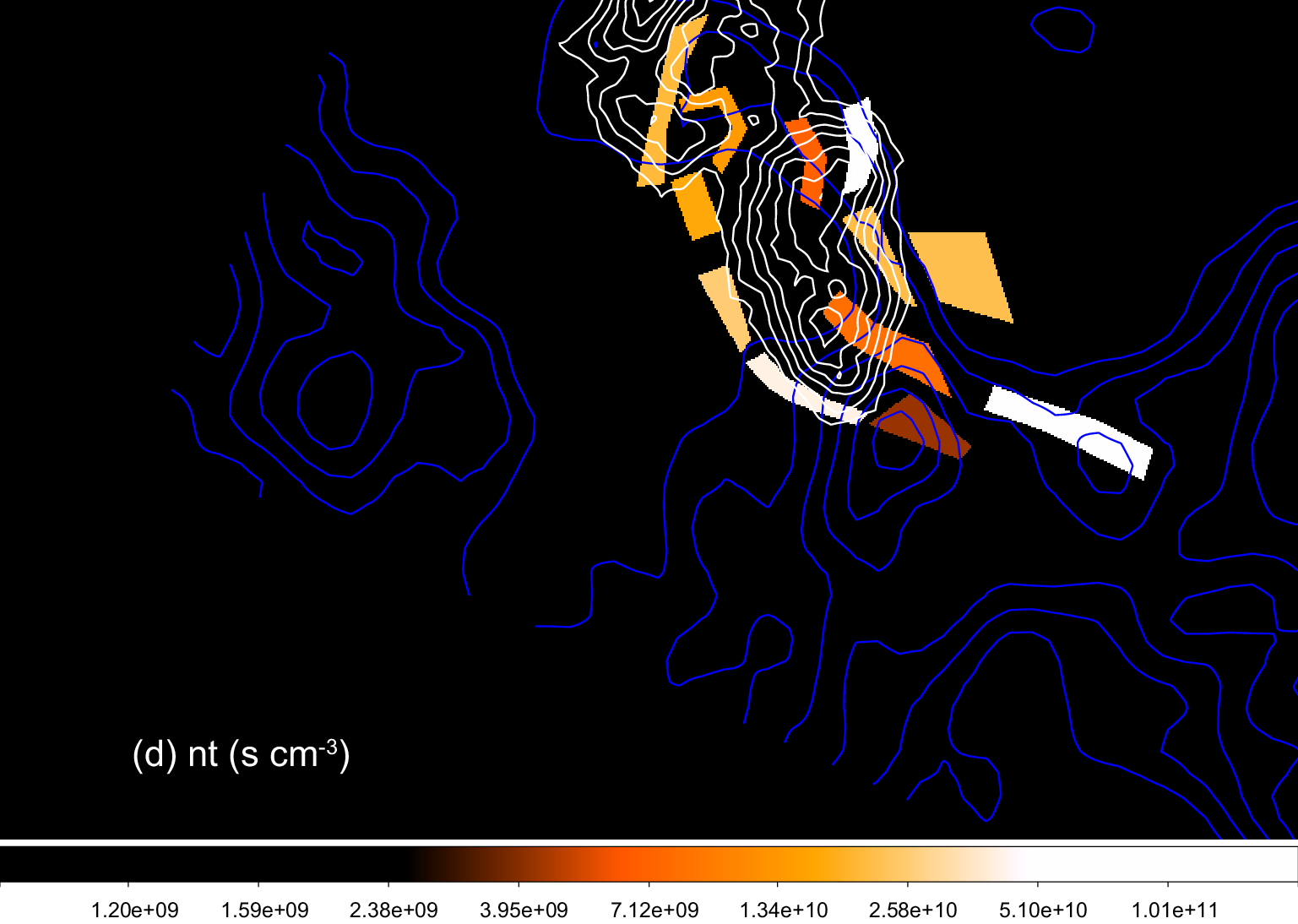}
\includegraphics[width=0.4\textwidth]{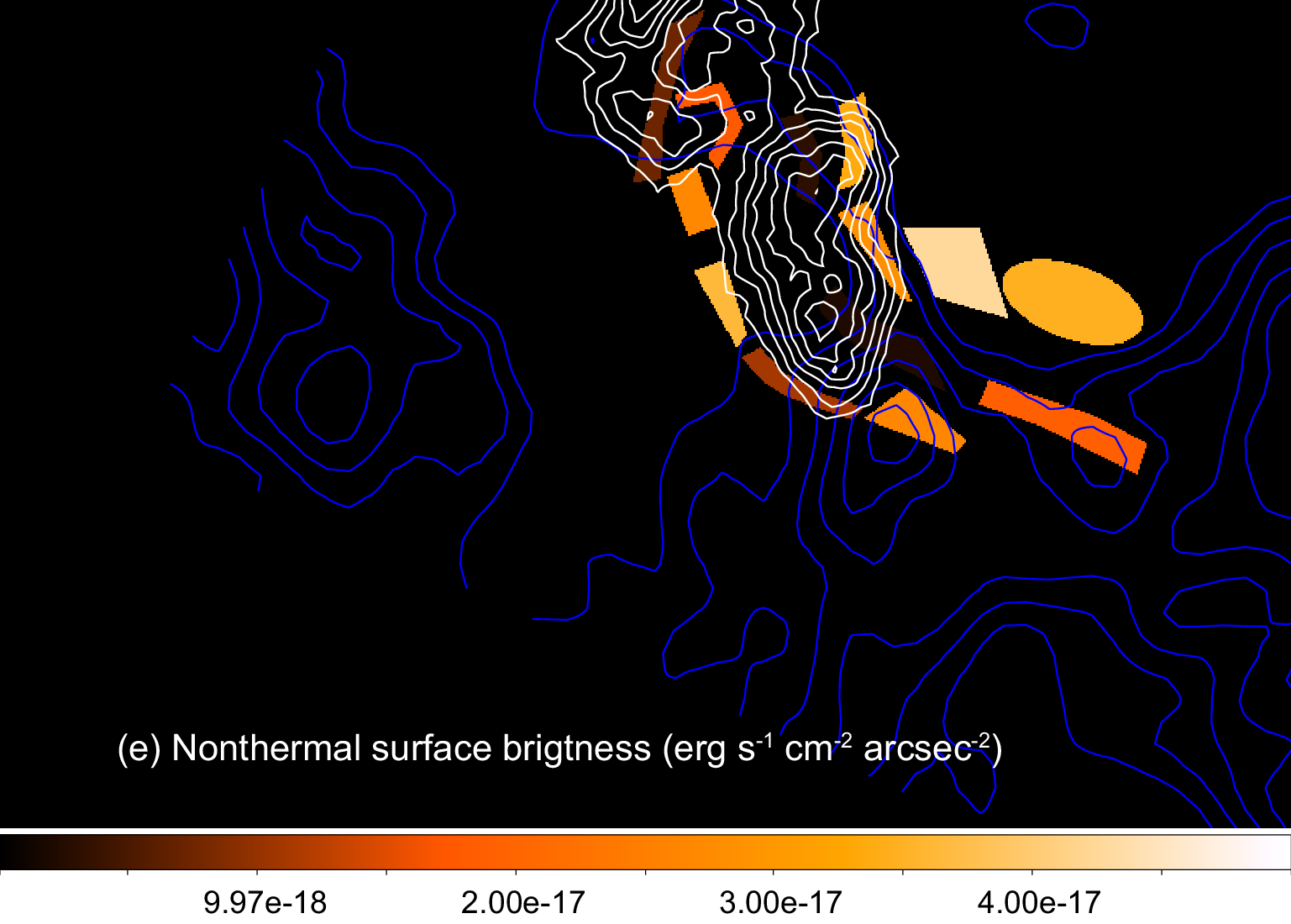}
\includegraphics[width=0.4\textwidth]{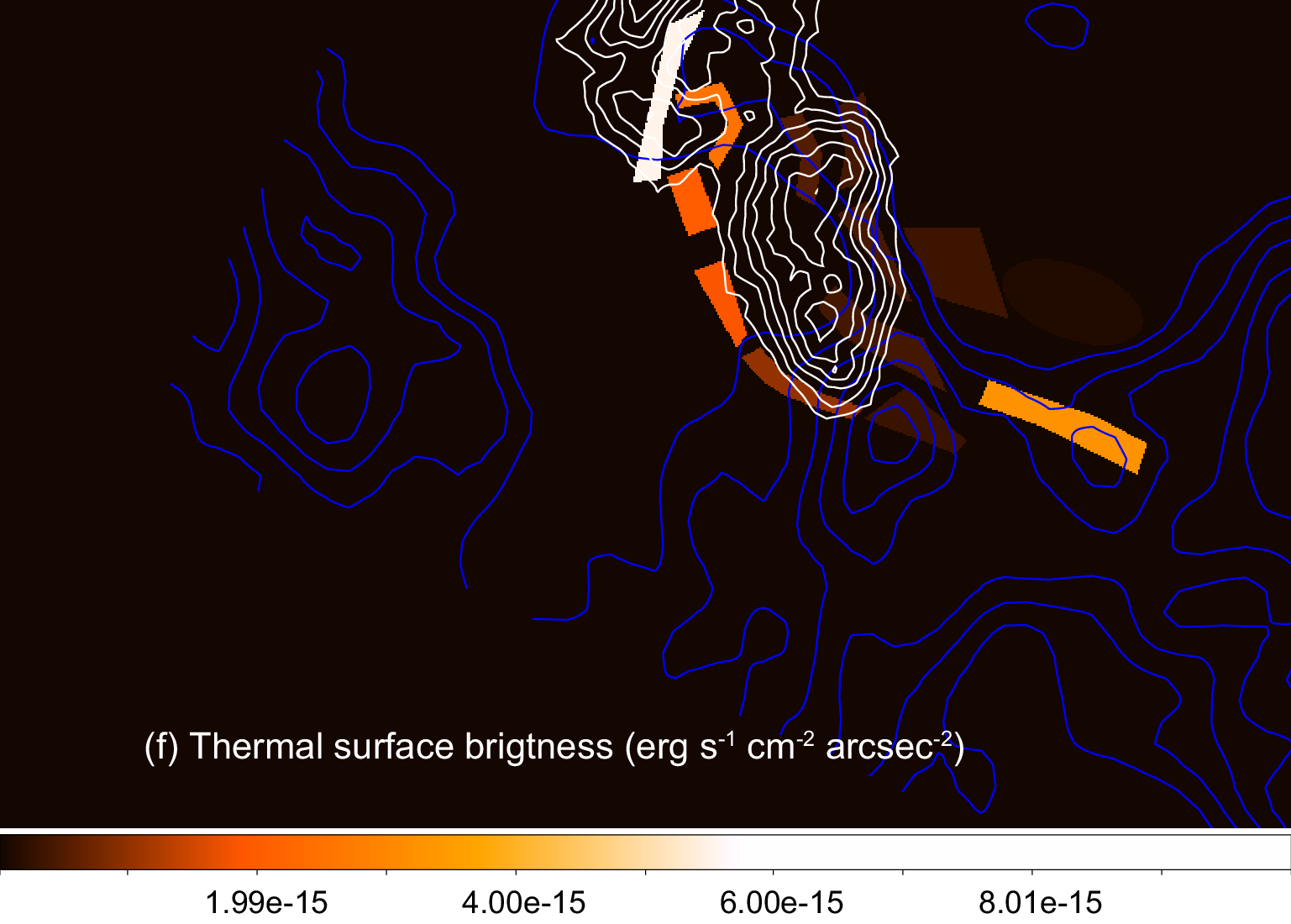}
\includegraphics[width=0.4\textwidth]{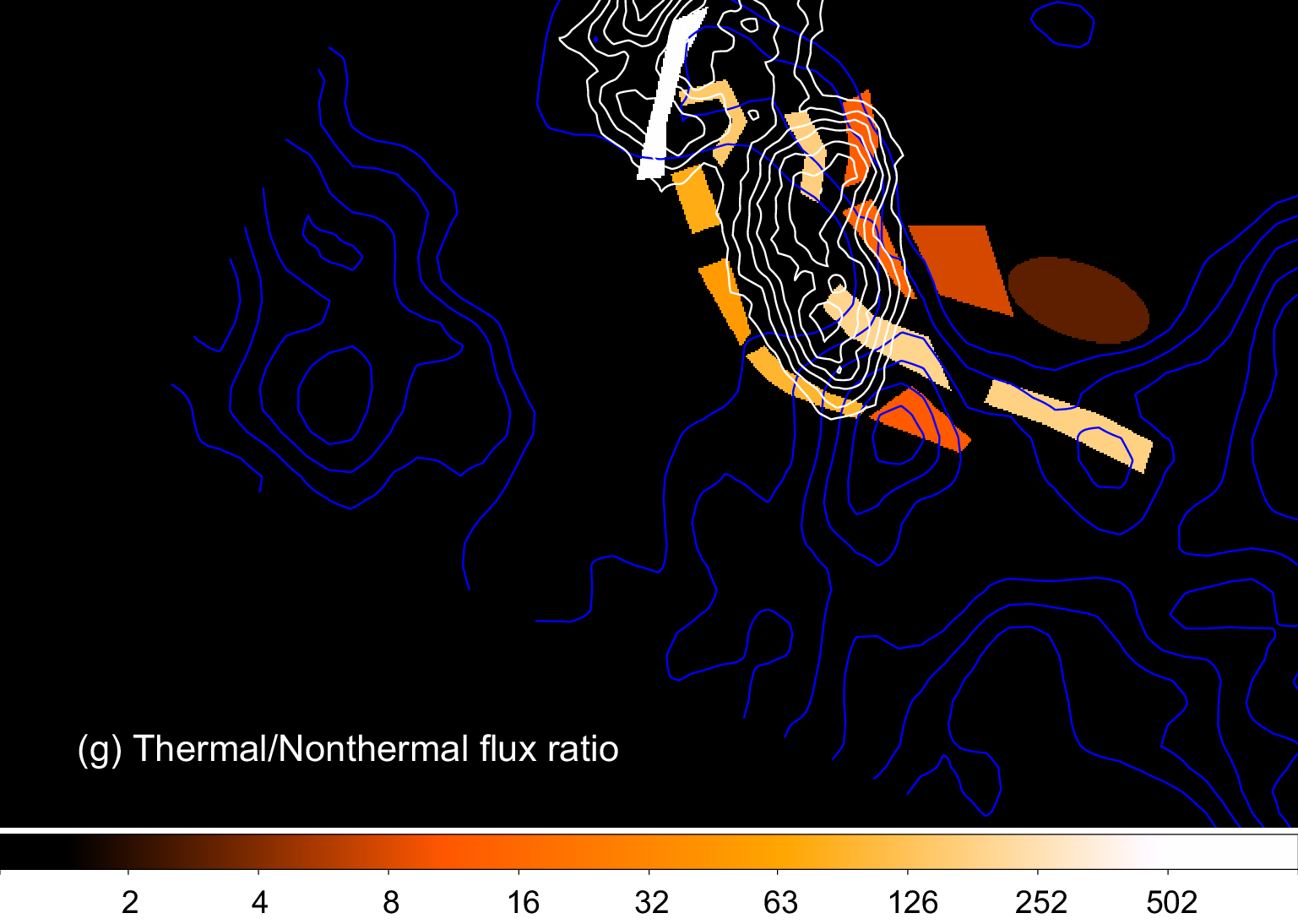}
 \end{center}
\caption{Parameter maps of the spectral fitting,
(a) absorption column,
(b) photon index,
(c) temperature,
(d) ionization timescale,
(e) nonthermal X-ray surface brightness in the 2--10~keV band in the unit of erg~s$^{-1}$cm$^{-2}$arcsec$^{-2}$,
(f) thermal X-ray surface brightness in the 0.5--10~keV band in the unit of erg~s$^{-1}$cm$^{-2}$arcsec$^{-2}$,
and 
(g) flux ratio between thermal and nonthermal emission.
Color scales are in linear for (a), (b), and (c), (e), and (f),
whereas in logarithmic for (d) and (g).
For (c) and (d), We have no data for src2 region.
White and blue contours represents
$^{12}$CO and H$_{\rm I}$ distribution, same as Fig~\ref{fig:image_CO} and Fig~\ref{fig:image_HI}, respectively.
}\label{fig:parameter_maps}
\end{figure}

\begin{figure}
\begin{center}
\includegraphics[width=0.4\textwidth]{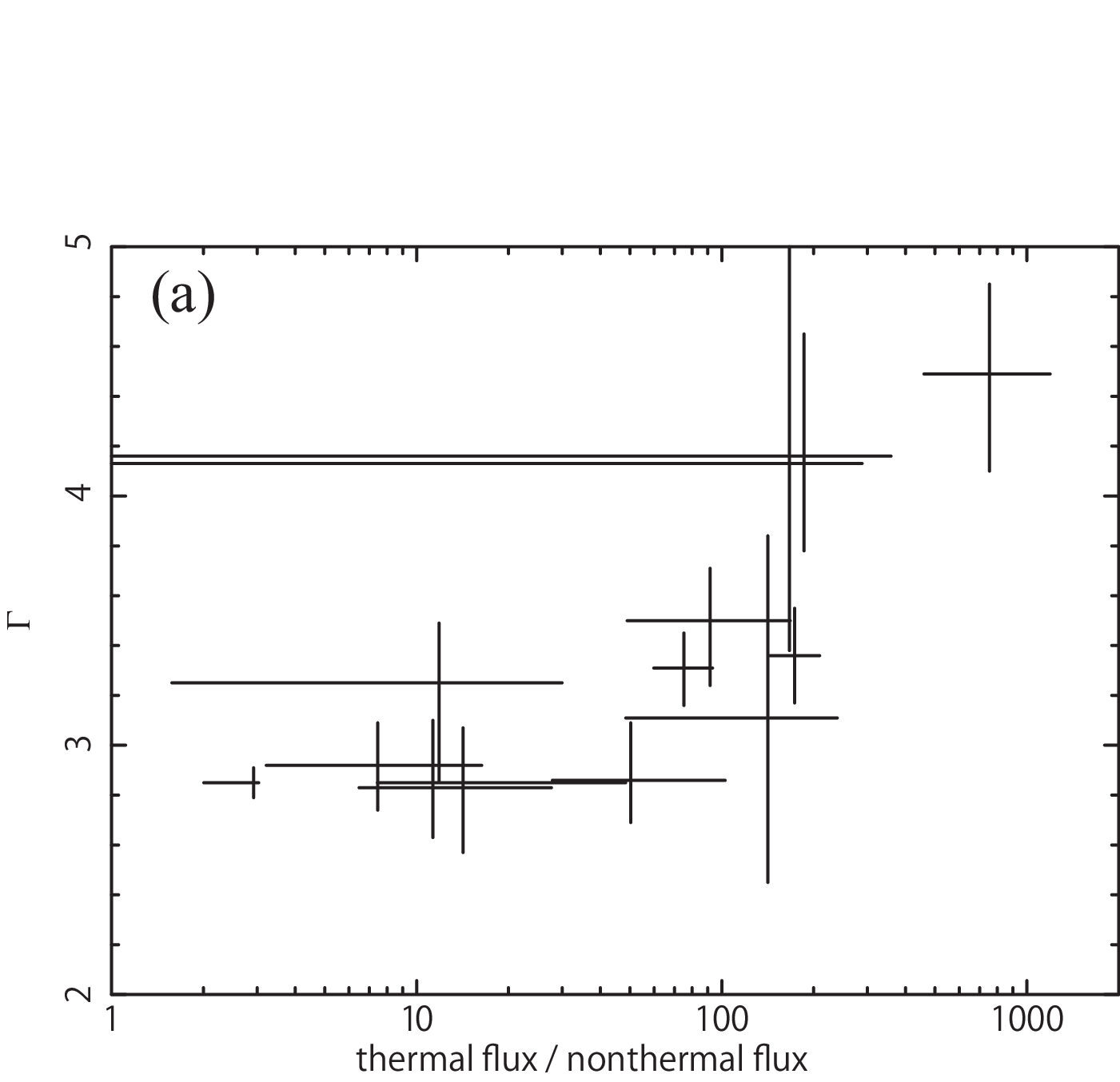}
\includegraphics[width=0.4\textwidth]{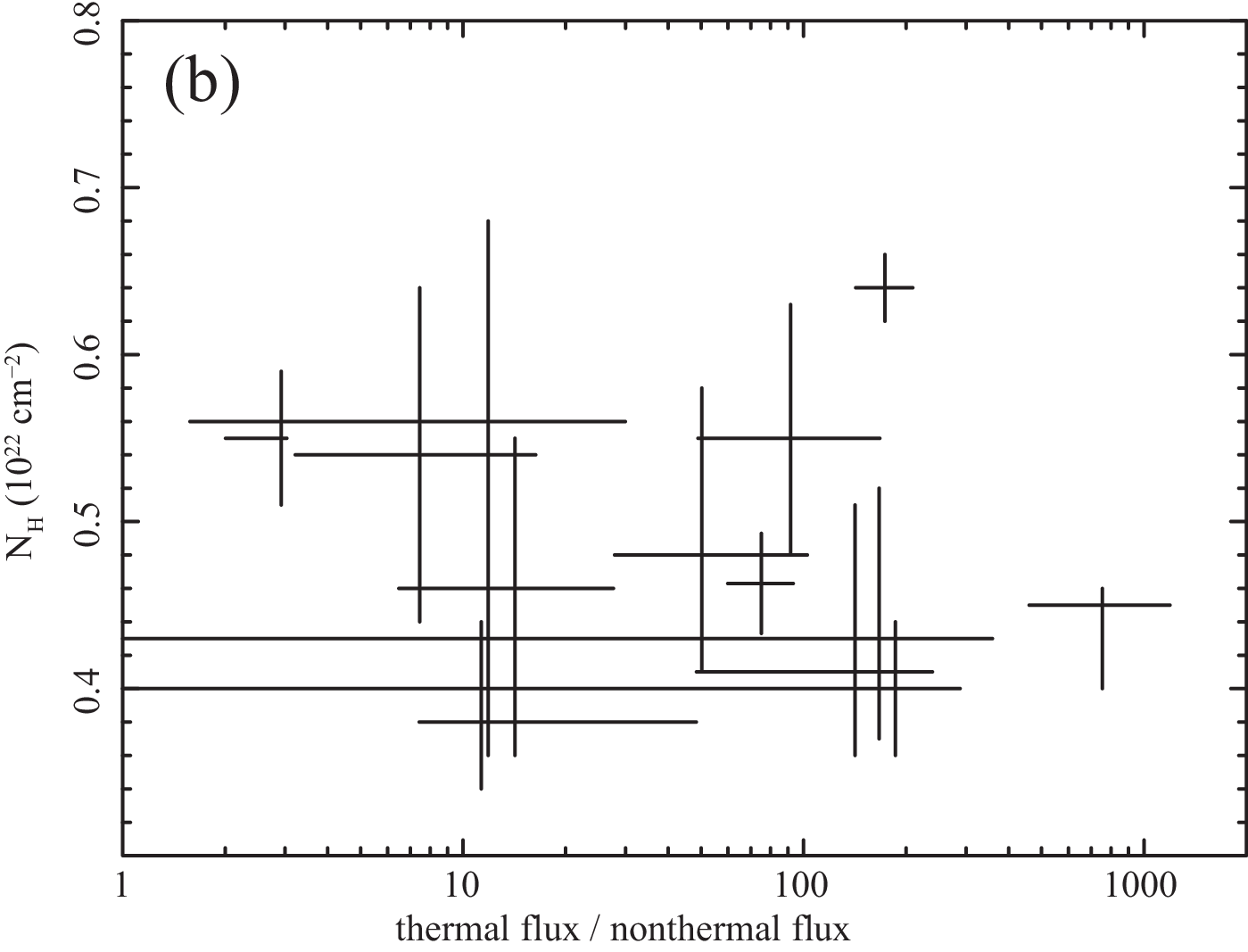}
\includegraphics[width=0.4\textwidth]{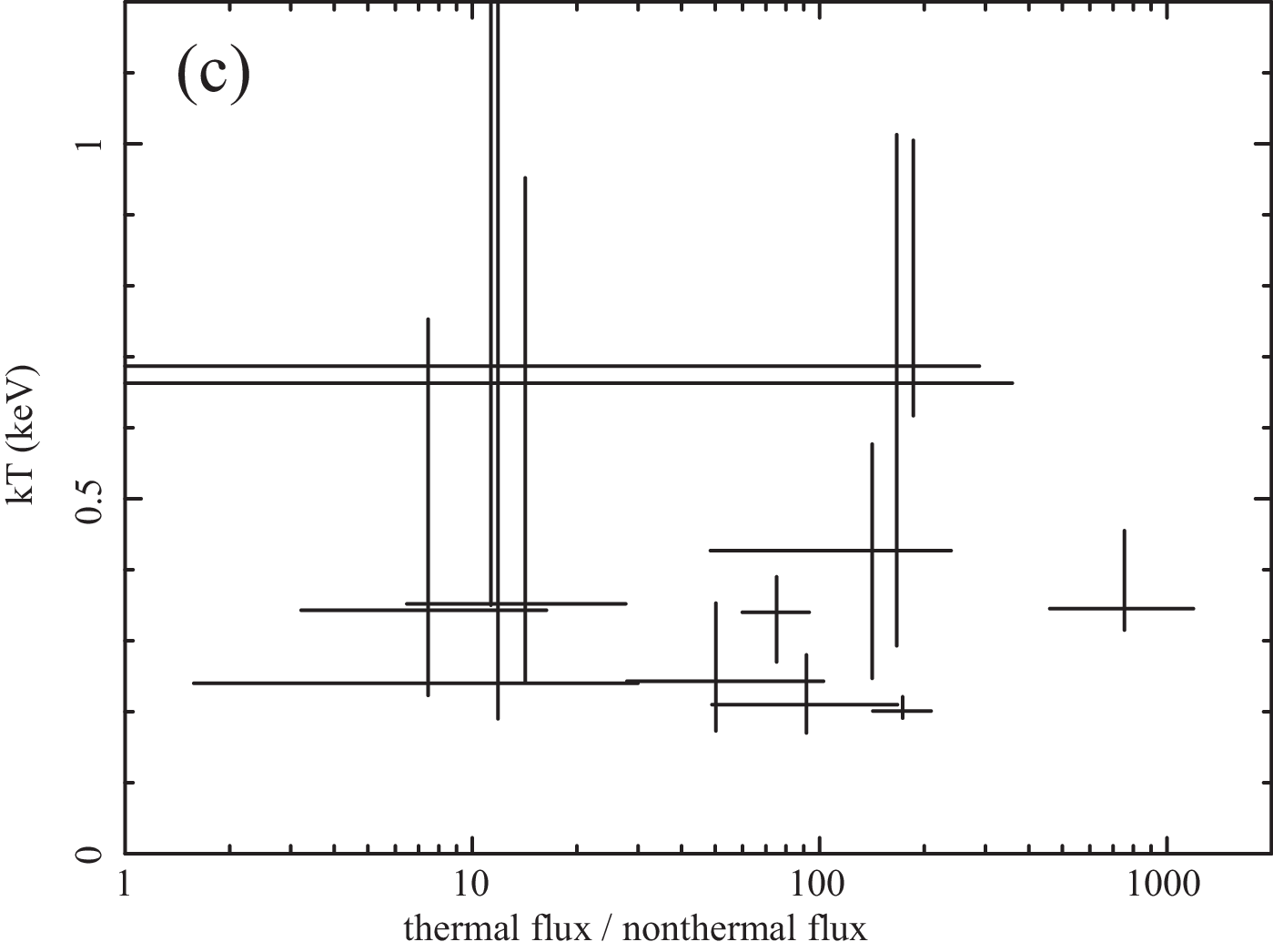}
\includegraphics[width=0.4\textwidth]{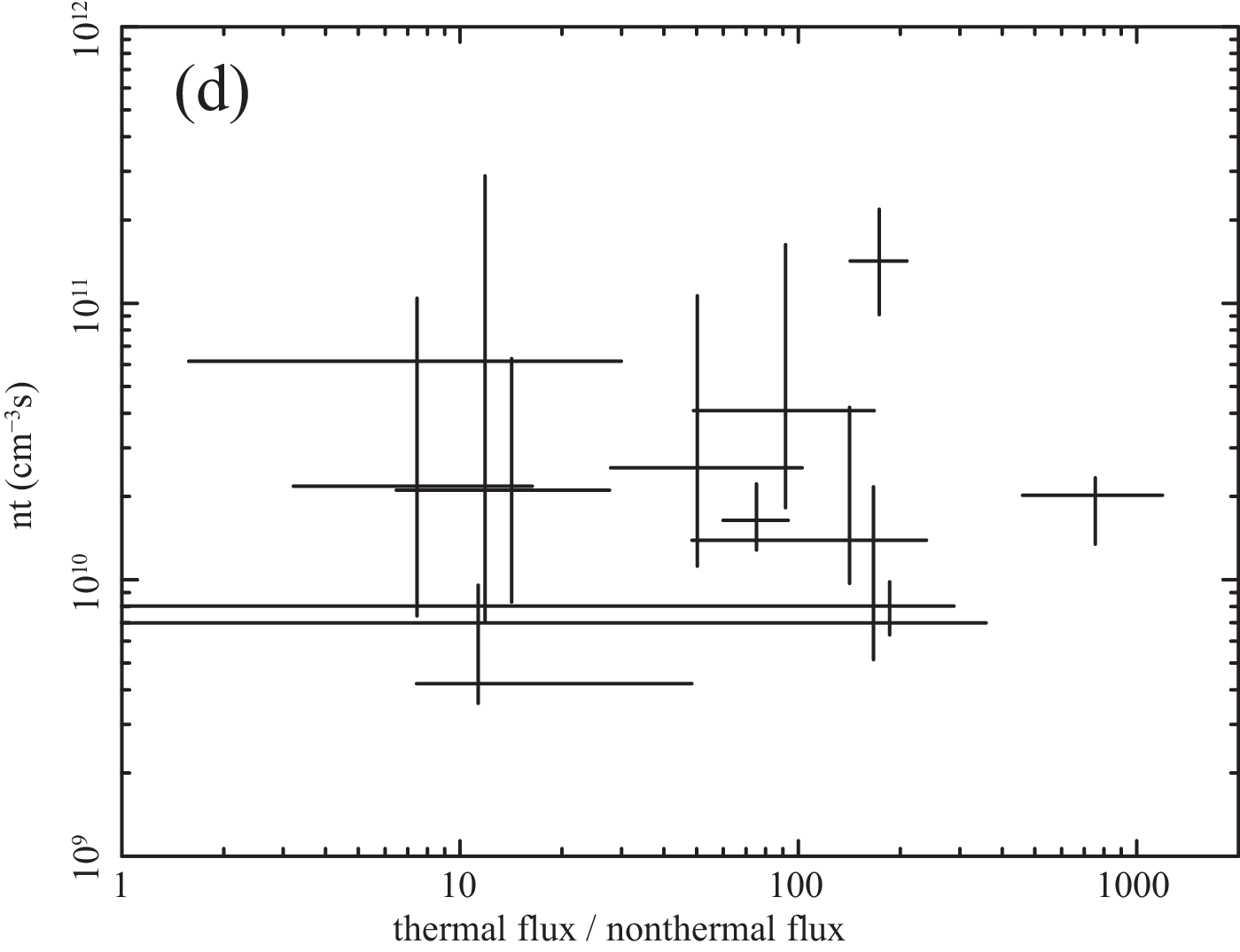}
\end{center}
\caption{Correlation plot between the flux ratio of the thermal and nonthermal components and the photon index of nonthermal emission (a), $N_{\rm H}$ (b), $kT_e$ (c), and $n_et$ (d).}\label{fig:correlation}
\end{figure}

Figure~\ref{fig:parameter_maps} show the parameter maps of the spectral analysis result.
The absorption column looks smaller
on the cloud regions (see Fig.~\ref{fig:parameter_maps}(a)), although
it is not so obvious.
Fig.~\ref{fig:parameter_maps}(b) shows that the nonthermal component has larger photon index on the shock-cloud interacting regions (src1, 3, 5, 6, 11, 12, 13).
These regions are bright in thermal X-rays (Fig.~\ref{fig:parameter_maps}(f)
but not so bright in nonthermal X-rays compared with non-interacting regions (Fig.\ref{fig:parameter_maps}(e).
Actually, the flux ratio between thermal and nonthermal emission changes two orders of magnitude as shown in Fig.\ref{fig:parameter_maps}(g),
implying that the interacting regions are really bright only in thermal X-rays.
To illustrate this further,
the panel (a) of Fig.~\ref{fig:correlation} shows
the scatter plot between the flux ratio and the photon index of nonthermal component.
One can see that
the photon index is larger---i.e. the nonthermal spectrum is softer---in the regions
where the thermal X-rays are dominant.
Ideally we would like to compare the molecular-cloud densities with the X-ray parameters directly.
However, the X-ray emission enhancement is on the edge of the molecular cloud, not on the center, which is difficult to see in the correlation plot. It is also unknown whether
the cloud is physically interacts with the shock or not.

On the other hand, the thermal parameters, $kT$ (Fig.~\ref{fig:parameter_maps}(c)) and $nt$ (Fig.~\ref{fig:parameter_maps}(d)), show no clear tendency on the correlation with dense clouds.
The correlation plot shown in the panel (b)--(d) of Fig.~\ref{fig:correlation} also support no clear correlation.

\section{Discussion}
\label{sec:discuss}

We carried out the spatially resolved spectroscopy
and found that 
filaments interacting with the dense cloud have
enhanced thermal emission, and fainter and possibly softer synchrotron X-rays.
The absorption column on the cloud regions is rather small,
which may imply that the cloud is on the rear side of the remnant.

The 
apparent
positive correlation between the flux ratio of thermal and nonthermal X-rays and the photon index of nonthermal X-rays shown in Figure~\ref{fig:correlation}(a) can be mainly due to 
a deceleration of the shock velocity caused by the interaction of the shock with dense material.
In the synchrotron loss-limit regime, the roll-off energy of the synchrotron X-rays is proportional to the square of shock velocity \citep{aharonian1999,yamazaki2006,zirakashvili2007}.
Thus the interacting shocks should have smaller shock velocity,
and softer (and as a result fainter) synchrotron X-rays.
The thermal X-rays are enhanced by the shock compression,
which results in bright soft X-ray filaments around the clouds.
We thus conclude that
this region of RCW~86 well follows the synchrotron loss-limit regime.

The enhancement of thermal X-rays, fainter and softer synchrotron X-ray emission is also shown 
in the interacting regions of several SNRs,
such as the northwestern rim \citep{sano2022,bamba2008}
of SN~1006
and Tycho \citep{lopez2015}.
In most SNRs bright  X-ray synchrotron emission appears also to coincide with no
or  very faint thermal X-ray emission (for example, \cite{nakamura2012}).
These facts all imply that 
the bright synchrotron X-rays are emitted around
non-interacting, fast shocks.
Our results on RCW86 shows such a thermal X-ray enhancement and decline of synchrotron X-rays
in the interacting regions with smaller spatial scales, 1~arcmin or 0.7~pc at 2.3~kpc distance, compared with previous results with the spatial scales of larger than a few pc.

On the other hand, our result is in contrast to the case of RX~J1713.7$-$3946 \citep{sano2013},
where synchrotron X-rays are enhanced around the shock-cloud interacting regions.
Similarity to RX~J1713$-$3946 is also observed in Kepler;
\citet{sapienza2022} revealed that the acceleration efficiency in the Kepler is higher in the north, interacting region with the shock and circumsteller medium, compared with the south of the remnant.
Even in the same remnant, RCW~86,
\citet{suzuki2022} showed that thermal and synchrotron X-rays
are enhanced in the denser region.
\citet{inoue2012} suggest that such a tendency happens on shocks colliding with clumpy clouds;
when the shock runs into clumpy medium,
the dense clumps can survive in the shock downstream region without evaporation
and amplify the downstream magnetic field,
which makes enhanced synchrotron X-rays
(see also \cite{sano2021}).

This scenario may indicate that
the difference between the shock-cloud interacting region in RCW86 and RX~J1713.7$-$3946
can be the difference of environment of the shock.
In RX~J1713.7$-$3946 case, the inter-cloud density is $\sim$ 0.01--0.1$~{\rm cm}^{-3}$ \citep{weaver1977,cassamchenai2004,katsuda2015}.
\citet{sano2020} discovered shocked cloudlets
with the density of $\sim 10^4 {\rm cm}^{-3}$,
thus the clumpiness, the density contrast between dense cloud and inter-cloud regions, should be high. 
This is consistent with the calculation by \citet{celli2019}.
On the other hand, the density contrast between clouds and inter-cloud medium surrounding RCW86's eastern region  can be smaller,
since the dense cloud detected in $^{12}$CO is covered by less-dense inter-cloud, which is detected with H$_{\rm I}$ emission.
The cloud interacting with SN~1006 is also detected by H$_{\rm I}$ observation only \citep{sano2022},
implying a less clumpy interstellar medium.
Similar discussion is also done in \citet{sano2021}.
On the interaction between the shock and such dense and rather uniform clouds,
the shock-heated plasma can emit strong thermal bremsstrahlung,
which should be the bright thermal filaments we observed.
In order to test our scenario, we need high spatial resolution CO and H$_{\rm I}$ observations of RCW86
to compare southwest (RX~J1713-like) and eastern (SN~1006 like) regions to see if there are high density clouds and if density contrast makes the synchrotron X-ray enhancement.

In this study,
we found no significant variation of temperature and ionization time scale of thermal emission.
This is mainly due to the lack of energy resolution and statistics
which lead large error ranges.
Future X-ray missions with excellent energy resolution,
such as XRISM \citep{tashiro2020} and Athena \citep{nandra2013},
will resolve this issue.

\section{Summary}
\label{sec:summary}

The spatially resolved spectroscopy of the southeastern region of the young supernova remnant RCW86 is done with deep XMM-Newton observation.
It is found that soft thermal X-rays are enhanced on the edge of dense clouds which is detected with $^{12}$CO or H$_{\rm I}$ observations.
These regions show fainter and possibly softer nonthermal X-rays from accelerated electrons.
These results indicate that
the shock decelerate due to the interaction with clouds, which makes 
thermal X-ray enhancement due to the density increase and fainter and softer synchrotron X-rays due to smaller shock speed.
Our results indicate that the dense region does not enhance synchrotron X-rays, in agreement with the suggestions by
\citet{aharonian1999,yamazaki2006,zirakashvili2007}.
No synchrotron X-ray enhancement on the shock-cloud interaction region is found in this region.
This is similar to SN~1006 and Tycho, and contrastive to 
RX~J1713.7$-$3946.
This difference can be due to the difference of
the surrrounding environment,
such as high density contrast of surrounding material.
For further study of the effect of surrounding material
to the synchrotron X-ray enhancement,
we need follow-up observations with ALMA to study
the density contrast of surrounding material of RCW86,
and XRISM/Athena observations to measure precise parameters of thermal emission to understand the thermal conditions.

\begin{ack}
We thank the anonymous referee for his/her productive comments.
We thank Hiromichi~Okon and Yukikatsu Terada who helped us to make the parameter maps.
This work was financially supported by Japan Society for the Promotion of Science Grants-in-Aid for Scientific Research (KAKENHI) Grant Numbers, JP19K03908 (AB), JP23H01211 (AB), 20KK0309 (HS), 21H01136 (HS), 22H01251 (RY), and 23H04899 (RY).
\end{ack}



\end{document}